\documentclass[aps,prb,twocolumn,superscriptaddress]{revtex4-2}

\usepackage{graphicx}
\usepackage{dcolumn}
\usepackage{bm}

\usepackage{graphicx}
\usepackage{derivative}
\usepackage{dcolumn}
\usepackage{bm}
\usepackage{hyperref}
\usepackage{braket}
\usepackage{amsmath}

\usepackage{amsfonts}

\begin{document}

\preprint{APS/123-QED}

\title{Subsystem localization in a two-leg ladder system}

\author{Arpita Goswami}
\thanks{These authors contributed equally}
\author{Pallabi Chatterjee}
\thanks{These authors contributed equally}
\author{Ranjan Modak}
\thanks{Contact author: ranjan@iittp.ac.in}
\author{Shaon Sahoo}
\thanks{Contact author: shaon@iittp.ac.in}
\affiliation{Department of Physics, Indian Institute of Technology Tirupati, India, 517619}

\date{\today}

\begin{abstract}
We consider a ladder system where one leg, referred to as the ``bath", is governed by an Aubry-Andr\'{e} (AA) type Hamiltonian, while the other leg, termed the ``subsystem", follows a standard tight-binding Hamiltonian. We investigate the localization properties in the subsystem induced by its coupling to the bath. For the coupling strength larger than a critical value ($t'>t'_c$), the analysis of the static properties shows that there are three distinct phases as the AA potential strength $V$ is varied: a fully delocalized phase at low $V$, a localized phase at intermediate $V$, and a weakly delocalized (fractal) phase at large $V$. The fractal phase also appears in a narrow region along the boundary between the delocalized and localized phases.
An analysis of the projected wavepacket dynamics in the subsystem shows that the delocalized phase exhibits a ballistic behavior, whereas the weakly delocalized phase is subdiffusive. Interestingly, the narrow fractal phase shows a super- to subdiffusive behavior as we go from the delocalized to localized phase.
When $t'<t'_c$, the intermediate localized phase disappears, and we find a delocalized (ballistic) phase at low $V$ and a weakly delocalized (subdiffusive) phase at large $V$. Between those two phases, there is also an anomalous crossover regime where the system can be super- or subdiffusive. Beyond the ballistic phase observed at low $V$, we also identify a superdiffusive regime emerging in the limit $t'/V \ll 1$, which continuously approaches the ballistic behavior as $t' \to 0$.
Finally, in some limiting scenario,  we also establish a mapping between our ladder system and a well-studied one-dimensional generalized Aubry-Andr\'{e} (GAA) model.

\end{abstract}

\maketitle


\section{Introduction}
Anderson localization is a key quantum effect where uncorrelated site disorder localizes wavefunctions and blocks transport of non-interacting electrons~\cite{anderson.1958,go4.1979,rmp_AL_1985}. It creates in a three dimensional system a mobility edge that separates localized and extended states. A metal-insulator transition occurs when the Fermi energy crosses this edge. In one dimension (1D), all states become localized with any amount of disorder. There is no mobility edge and no transition. Introducing interactions in such disordered Anderson localized systems can stabilize a many-body localized (MBL) phase, where systems fail to thermalize and retain memory of initial conditions~\cite{MBL1,MBL_rev1,MBL2,MBL3,MBL4,basko2006metal,nandkishore2015many,ghosh2019many}. This behavior contradicts the Eigenstate Thermalization Hypothesis (ETH), which posits that isolated non-integrable systems should relax to thermal equilibrium~\cite{ETH1,ETH2,rigol2012alternatives,rigol2009breakdown}. However, recent studies highlight the avalanche instability within the MBL phase. This instability suggests that rare regions of the system can thermalize and the growth of entanglement between different such regions potentially destabilizing the MBL phase~\cite{MBL_breakdown1,MBL_breakdown2,vidmar.2020}.

On the other hand, the presence of correlated disorder can significantly change the Anderson localization scenario in 1D. 
Such systems may have a localization-delocalization transition and even may have a mobility edge.
One of the prime examples of such systems is the  Aubry-Andr\'{e} (AA) Hamiltonian, where the uncorrelated ``true" disorder is replaced by a quasi-periodic potential ~\cite{aubry1980analyticity}. One can induce a delocalization–localization transition by tuning the strength of the quasi-periodic potential, a phenomenon that holds significance in various physical contexts ~\cite{aa_addi2,aa_addi1,aa_addi3,aa_addi4,aa_addi5}.  The AA Hamiltonian also has the unique property of  self-duality at the critical point, with identical real space and momentum space representations, thus it contains an energy-independent
localization transition for the whole spectrum with no
mobility edges. It has been shown that by adding different types of perturbations to the AA model, the fine-tuned AA duality conditions can be broken, and an energy-dependent self-duality relation can be established, automatically implying the existence of the mobility edges~\cite{spme_1,spme_2,spme_3,spme_4,spme_5,spme_6,analytic_spme1,analytic_spme2}.  With the remarkable progress in cold-atom and ion-trap experiments over the past decade, many of these models have been experimentally realized~\cite{exp1,exp2,exp3,exp4}, providing strong motivation to study quasi-periodic systems beyond purely theoretical interest. In this context, the emergence of a non-ergodic metallic phase has been proposed in models with single-particle mobility edges when interactions are introduced~\cite{nem_modak,nem_sriram,deng2017many,nem_soumi,garg.17}.  

A question that has persisted for some time now is what happens if one couples localized and extended systems together? In the case of a many-body interacting system, if the degrees of freedom of the extended system are large enough compared to the localized system, the extended system is expected to thermalize the localized one. On the other hand, if the delocalized degrees of freedom are comparable to the localized ones, the outcome is not yet clear. Many efforts have been made to address this question in the last several years~\cite{dlvslcoupling_1,dlvslcoupling_2,dlvslcoupling_3,dlvslcoupling_4,dlvslcoupling_5}. Numerical challenges to tackle the exponentially growing Hilbert space dimension with system size and limitations of analytical tools to solve interacting systems have made this problem extremely challenging.   
In this work, we address a similar question in the non-interacting context.

\begin{figure}[t]
\includegraphics[width=1\linewidth]{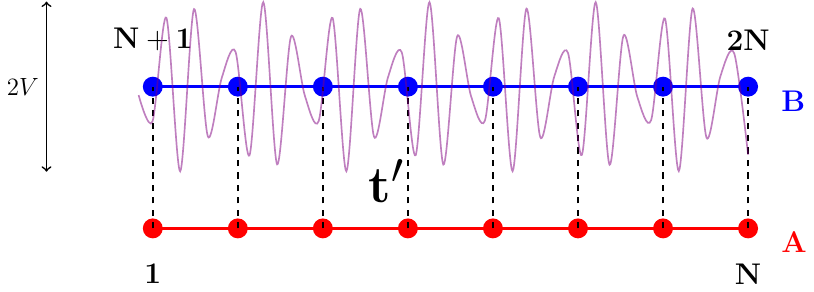}
    \caption{Schematic diagram of the model: the subsystem ($A$) is described by a standard TB Hamiltonian whereas the bath ($B$) is governed by an AA type potential.}   \label{fig:system}
\end{figure}

We consider a ladder system where one leg, referred to as the “bath" is a 1D non-interacting disordered chain, while the other leg, termed the “subsystem”, is described by a clean delocalized tight-binding (TB) Hamiltonian (see the schematic of our system in Fig.~\ref{fig:system}).  If the disorder in bath is uncorrelated, according to the rule of Anderson localization (i.e., any tiny disorder is sufficient to localize all states in 1D and 2D), both the subsystem  and bath of the total system are expected to be Anderson localized. However, if the disorder is correlated, the outcome is not at all apparent, given that the so-called rule of Anderson localization does not apply there. We investigate such a scenario by introducing a quasi-periodic potential (AA type) in the bath. 
We find that, depending on the strength of the quasi-periodic potential $V$ and the coupling strength $t'$ between subsystem  and bath, it is possible to localize the subsystem.  One might expect such a situation to arise (if at all) in the large $V$ limit, when the bath is strongly localized. Interestingly, it turns out that is not the case. In contrast, this situation arises when $t'$ is larger than a critical strength $t'_c$, and that too 
for an intermediate range of $V$. In the limit of large $V$, the subsystem remains weakly delocalized and exhibits subdiffusive transport. Finally, in some limiting scenario, we
also establish a mapping between our system and a well-studied one-dimensional generalized Aubry-Andr\'{e} (GAA) model, which also 
exhibits similar ballistic, superdiffusive, and subdiffusive
phases, as reported recently~\cite{dynamics_2}.

Manuscript is organized as follows. In Sec.~\ref{sec:model}, we discuss our model and outline the main outcome in the form of a phase diagram. In Sec .~\ref{sec3}, we investigate the static properties of the subsystem for the significant states. Sec.~\ref {sec4} is dedicated to studying the projected wave-packet dynamics in the subsystem. In Sec.~\ref {sec:analytic}, we discuss the analytical understanding of our results.  Finally, in Sec.\ref{sec:conclusions}, we briefly summarize our main conclusions.

\section{Model system and phase diagram}\label{sec:model}
In this section, we first describe the model system that we investigate. Next, we provide a summary of our results and present a schematic phase diagram.  

\subsection{Model system}
In this paper, we investigate a ladder system whose one leg is called the ``subsystem" and the other leg is called the ``bath". These are, respectively, indicated by ``$A$" and ``$B$" in the schematic diagram (Fig. \ref{fig:system}). The subsystem $A$ is described by a standard tight-binding (TB) Hamiltonian ($H_A$ in Eq. \ref{ham_eqn}) while the bath is governed by an Aubry-Andr\'{e} (AA) type Hamiltonian ($H_B$ in Eq. \ref{ham_eqn}). Without the coupling ($t'=0$) between the subsystem $A$ and bath $B$, all the states in $A$ are delocalized. The problem we investigate here is whether and when states become localized in $A$ as we establish a coupling ($t'\ne0$) between $A$ and $B$. 

\begin{figure}[t]
\includegraphics[width=0.9\linewidth]{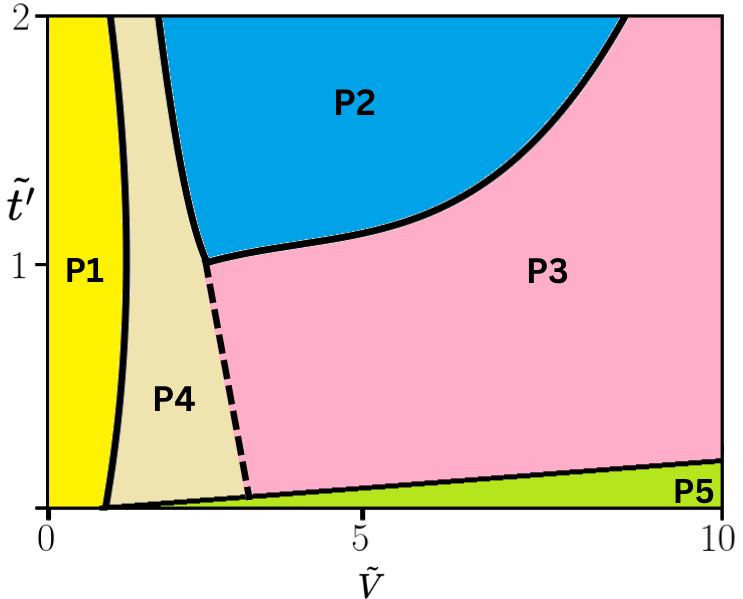}
    \caption{Schematic phase diagram in rescaled parameter space of $\tilde{t'}=t'/t'_c$ and $\tilde{V}=V/t'_c$.    
    P1 - delocalized (ballistic) phase, P2 - localized phase, P3 - weakly delocalized (fractal and subdiffusive) phase, P4 - an anomalous crossover regime (fractal and super- or sub-diffusive), and P5 - weakly delocalized (fractal and superdiffusive) phase.}
    \label{fig:phases}
\end{figure}

The full Hamiltonian of the ladder system appears in the following equation:
\begin{equation}\label{ham_eqn}
\begin{split}
H &= H_A+H_B+H_{AB},~ \text{where}\\
H_A &= -t_A\sum_{j=1}^{N-1}(c_j^{\dagger}c_{j+1}+hc), \\
H_B &= -t_B\sum_{i=N+1}^{2N-1}(c_i^{\dagger}c_{i+1}+hc)\\
    &~~~~+ V\sum_{i=N+1}^{2N}\cos{(2\pi\beta i+\phi)}~c_i^{\dagger}c_i,~ \text{and}\\
H_{AB} &= -t'_{AB}\sum_{j=1}^N(c_j^{\dagger}c_{N+j}+hc).
\end{split}
\end{equation}
We take $t_A=t_B=1$ and $\beta=\frac{\sqrt{5}-1}{2}$. For our numerical study, we take $V$ and $t'_{AB}$ as two parameters. The second parameter $t'_{AB}$ will be denoted by $t'$ in the remainder of the paper. We do $\phi$ averaging for all our results for better statistics. 
\begin{figure*}[t]
    \centering
    \includegraphics[width=1\linewidth]{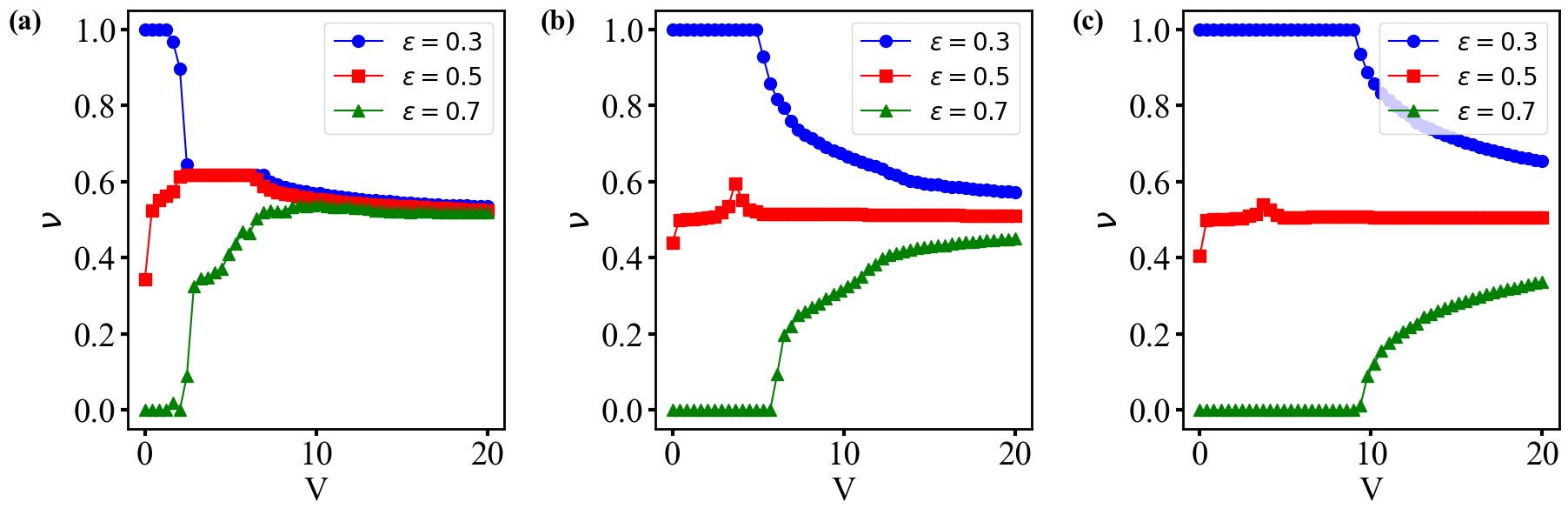}
    \caption{(a)-(c): $\nu$ vs. $V$ plots for $t'=$ 1, 5 and 10,  respectively. The calculations are performed for the total subsystem size N = 500.}
    \label{fig:nu_V}
\end{figure*}

\subsection{Phase Diagram}
The summary of our main results is presented in the schematic diagram of Fig. \ref{fig:phases}. 
It is constructed based on 
two scaling parameters: $\eta$ - finite-size scaling exponent of the average subsystem PR (Participation Ratio) and $\gamma$ - dynamical exponent in the long-time behavior of the MSD (Mean-Squared Displacement).
The analysis of the static properties, as discussed in Sec. \ref{sec3}, reveals that above a certain threshold value of the coupling strength ($t'>t'_c\approx 4.4$), we have three distinct phases - a fully delocalized phase (for small values of $V$), a fully localized phase (for intermediate values of $V$) and a weakly delocalized (fractal) phase (for large values of $V$). We also see a narrow crossover region with fractal character along the line separating the delocalized and localized phases. 
The study of projected wavepacket dynamics in subsystem, as discussed in Sec. \ref{sec4}, reveals that the delocalized phase is ballistic (referred to as P1 phase) and the weakly delocalized phase is subdiffusive (referred to as P3 phase) in nature. Interestingly, the narrow crossover region (referred to as P4 phase) is found to exhibit both super- and sub-diffusive behavior as we go across the region.

In the regime where $t'<t'_c$, the study of static properties shows the absence of the localized phase. Here, one finds two phases - a delocalized phase for small values of $V$  and a weakly delocalized (fractal) phase for large values of $V$, which also shows ballistic and sub-diffusive transport, and hence, identified as once again P1 and P3, respectively. The study of wavepacket dynamics reveals that P4 phase also exists for $t'<t'_c$ in some intermediate regime of $V$, which is a fractal phase but can be super- or subdiffusive in nature. Moreover, we also identify an additional superdiffusive regime, referred to as the P5 phase. Such a phase exists in the regime $t'/V\ll 1$, and this phase becomes ballistic in nature in the strict limit $t'\to 0$.

The schematic phase diagram (Fig.~\ref{fig:phases}) summarizes all the main results of this work. For the phase diagram, we rescale the axes as $\tilde{t'}=t'/t'_c$ and $\tilde{V}=V/t'_c$, and restrict our regime of interest to $\tilde{t'}\in [0,2]$ and $\tilde{V}\in [0,10]$.
Since it is possible that, for a certain regime, a finite fraction of the significant states are localized and the rest are delocalized (weakly delocalized), it is important to have clarification on 
the meaning of the exponent $\eta$ in such a case and related phases in our energy-independent phase diagram. As we discuss later (see Sec. \ref{sec3} and Appendix \ref{ER_IPR}), in the thermodynamic limits, all the significant states are delocalized (localized) in the P1 (P2) phase. In the fractal phase (P3, P4, and P5 phases together), a finite fraction of significant states are delocalized (weakly delocalized) in general, and the rest are localized. The scaling exponent $\eta$ is effectively determined by the average behavior of the delocalized (weakly delocalized) states. In this case, where we have both types of states, the phase diagram is determined by the nature of non-localized states since the dynamical nature of the subsystem (characterized by the exponent $\gamma$) is mainly dependent on these states.

It may be noted that the phase diagram presented here strongly depends on the nature of the potential in the bath $B$. Instead of a correlated AA-type potential, if we take a random potential (for example, replace the uniform phase factor $\phi$ in Eq. \ref{ham_eqn} by a site-dependent uncorrelated phase factor $\phi_i$), the subsystem $A$ will always be in a localized phase, in accordance with the Anderson localization in a disordered one-dimensional system. The relevant result is presented in Fig. \ref{fig:randm_phs} of Appendix \ref{cos_random}. A similar model with true disorder was recently examined in Ref.~\cite{lin2024fate}, though there the authors focused on the entire system, unlike our analysis of the subsystem.

\section{Static properties} \label{sec3}
In this section, we study the localization properties of the subsystem $A$, as a function of the coupling strength $t'$ and the AA potential strength $V$. For this purpose, we calculate the Inverse Participation Ratio for the subsystem (denoted by $IPR_A$) and the average Participation Ratio for the subsystem (denoted by $\langle PR_{A} \rangle$). 
Before we discuss about these measures and corresponding results, we first note that all the eigenkets of the full Hamiltonian do not have the same degree of significance when studied from the subsystem $A$. We, therefore, begin our discussion with an analysis on the significant states.

\begin{figure*}[t]
    \centering
    \includegraphics[width=0.9\linewidth]{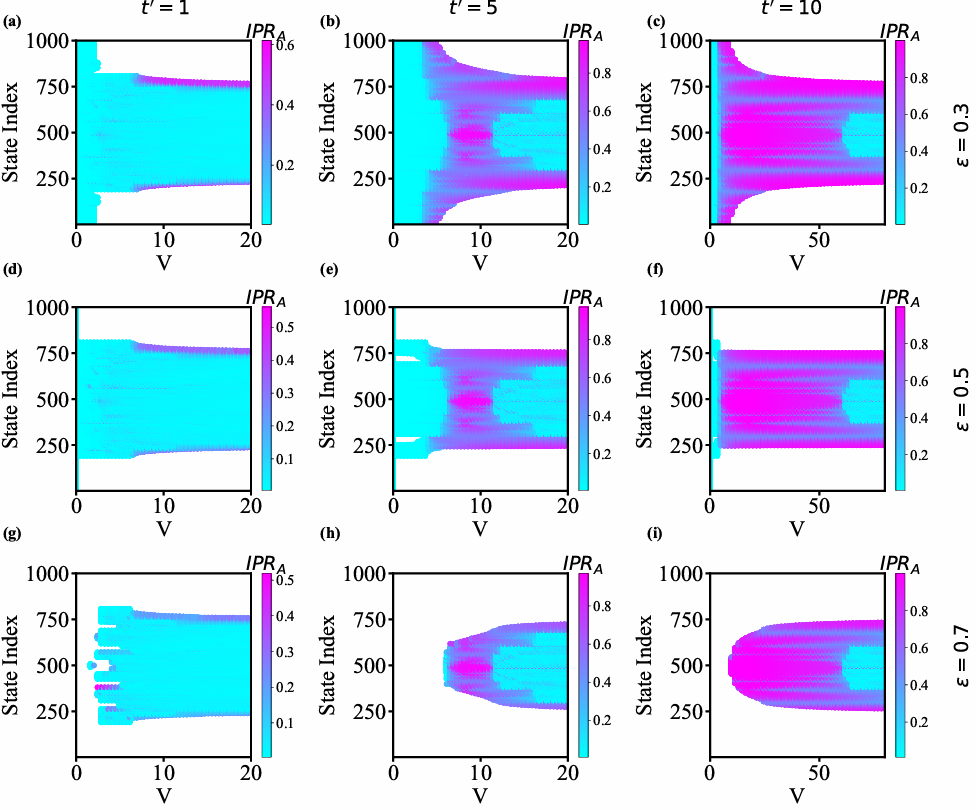}
    \caption{Plots of $IPR_A$ as function of $V$ for all $\epsilon$-significant states ($\epsilon$ = 0.3, 0.5 and 0.7). Calculations are performed for the total subsystem size N = 500 and for $t'$ = 1, 5 and 10.}
    \label{fig:IPR_ind_V}
\end{figure*}

\begin{figure}[!h]
    \centering
    \includegraphics[width=0.84\linewidth]{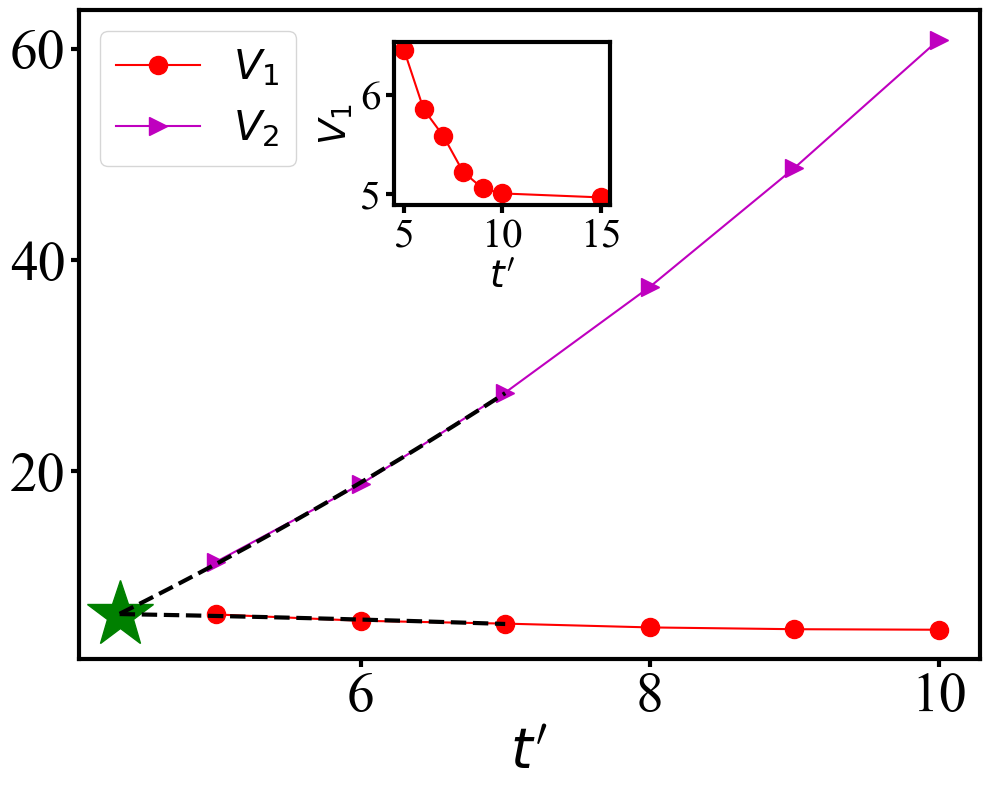}
    \caption{Variation of $V_1$ and $V_2$ as function of $t'$. The extrapolation of data shows that $V_1\approx V_2$ at $t'=t'_c\approx 4.4$ (represented by a star symbol).}
    \label{fig:V1_V2}
\end{figure}

\subsection{The $\epsilon$-significant states \label{sec3a}}
All the eigenkets of the full Hamiltonian do not have the same degree of significance when studied from the subsystem $A$. 
To study the localization on the subsystem $A$, we only analyze those eigenkets for which the subsystem probability $p_A$ is larger than certain value ($\epsilon$), where 
\begin{equation} \label{prob_A}
    p_A = \sum_{j \in A} |\braket{j|\xi}|^2,
\end{equation}
for an eigenket $\ket{\xi}$ of the full Hamiltonian. We choose the standard site basis $\{\ket{j}\}$ for our calculations, where $\ket{j}$ represents a ket corresponding to a single particle at the $j$th site (and no particles at any other site). The sum appearing in the equation is over all the sites in the subsystem $A$. An eigenket is called an {\it $\epsilon$-significant} if $p_A\ge\epsilon$. 
For $\epsilon =$ 0.3, 0.5 and 0.7, we calculate and analyze $IPR_A$ and $\langle PR_{A} \rangle$ (defined later) of the $\epsilon$-significant eigenkets.

To see how many eignkets are $\epsilon$-significant for the subsystem $A$, we define a parameter $\nu$ in the following way. Let $\mathcal{S}_{\epsilon}=\{ \ket{\xi}~ : p_A \ge \epsilon  \}$ be the set of $\epsilon$-significant eigenkets, i.e., the set of eigenkets for which the subsystem probability $p_A \ge \epsilon$. Since there are $2N$ number of eigenkets, we define:
\begin{equation} \label{nu_eps}
    \nu (\epsilon) = \frac{|\mathcal{S}_{\epsilon}|}{2N},
\end{equation}
where $|\mathcal{S}_{\epsilon}|$ is the number of elements in $\mathcal{S}_{\epsilon}$, and the argument in $\nu (\epsilon)$ explicitly shows that the parameter $\nu$ depends on $\epsilon$. The values of $\nu$ can be seen in Fig. \ref{fig:nu_V} for a wide range of $V$ and for three fixed values of $t'$. 
We get the following main results from this study. For a fixed value of $t'$, and for a small value of $V$, almost all eigenkets  are $\epsilon = 0.3$ significant (i.e., $\nu (0.3) = 1$) and almost no eigenket is $\epsilon = 0.7$ significant (i.e., $\nu (0.7) = 0$). As $V$ takes a value beyond a certain threshold (which depends on the given $t'$), we see that the number of the $\epsilon = 0.3$ significant states decreases (i.e. $\nu (0.3) < 1$), while the number of the $\epsilon = 0.7$ significant states grows (i.e. $\nu (0.7) > 0$). In the very large $V$ limit, $\nu \to 0.5$ for both the cases. Interestingly, for a wide range of $V$, we observe that $\nu (0.5)$ is always close to 0.5. This shows that, regardless of the values of $t'$ and $V$, there are about 50\% eigenkets that are $\epsilon = 0.5$ significant for the subsystem $A$, i.e., for about half of the eigenkets, the subsystem probability $p_A\ge 0.5$.    

We can understand the above results in the following way. For a fixed value of $t'$, and with a low value of $V$, most of the eigenstates are fully delocalized in the full system. As a result, for the individual eigenkets, $p_A\approx0.5$. Consequently, $\nu(0.3)\approx 1$, $\nu(0.5)\approx 0.5$ and $\nu(0.7)\approx 0$.  Now, as we increase $V$ to a large value, about half of the eigenkets (approximately $N$ in number) get localized in the bath $B$. This leaves about $N$ eigenkets that are localized or delocalized in the subsystem $A$. As a result, for large $V$, $\nu(0.3)\approx 0.5$, $\nu(0.5)\approx 0.5$ and $\nu(0.7)\approx 0.5$.

\subsection{$IPR_A$: Inverse Participation Ratio for subsystem}
To evaluate $IPR_A$, the inverse participation ratio for the subsystem $A$, we first project an eigenket of the full Hamiltonian (Eq. \ref{ham_eqn}) onto the subsystem $A$. We then calculate $IPR_A$ from the normalized projected state. If $\ket{\xi_n}$ is the $n$th eigenket of the full Hamiltonian, and $\hat{P}_A$ is the projector for the subsystem $A$, the corresponding normalized projected state for the subsystem is:
\begin{equation} \label{ket_A}
|\psi_n\rangle=\frac{\hat{P}_A|\xi_n\rangle}{\lVert\hat{P}_A|\xi_n\rangle\rVert}
= \frac{\sum_{j\in A} \langle j|\xi_n\rangle |j\rangle  } {\lVert \sum_{j \in A} \langle j|\xi_n\rangle |j\rangle\rVert}.
\end{equation} 
We define the subsystem $IPR_A$ in the following way: 
\begin{equation} \label{IPR_eq}
    \mathrm{IPR}_A^{(n)} = \sum_{j \in A} |\braket{j|\psi_n}|^4,
\end{equation} 
where $n$ is the eigenket index. 

\begin{figure*}[t]
    \centering
    \includegraphics[width=0.84\linewidth]{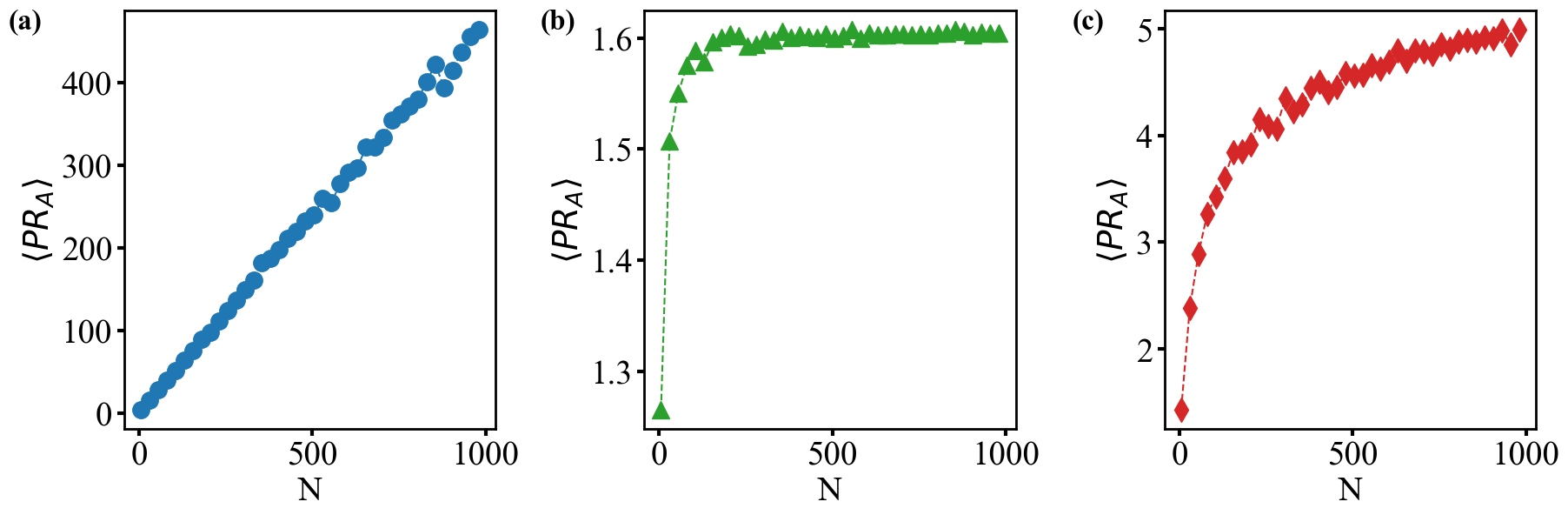}
    \caption{Plots of $\langle PR_{A} \rangle$ vs. N for $V=1,10,17$ ($\epsilon=0.5$ and $t'=5$).}
    \label{fig:PR_L}
\end{figure*}

\begin{figure}
    \centering
    \includegraphics[width=0.84\linewidth]{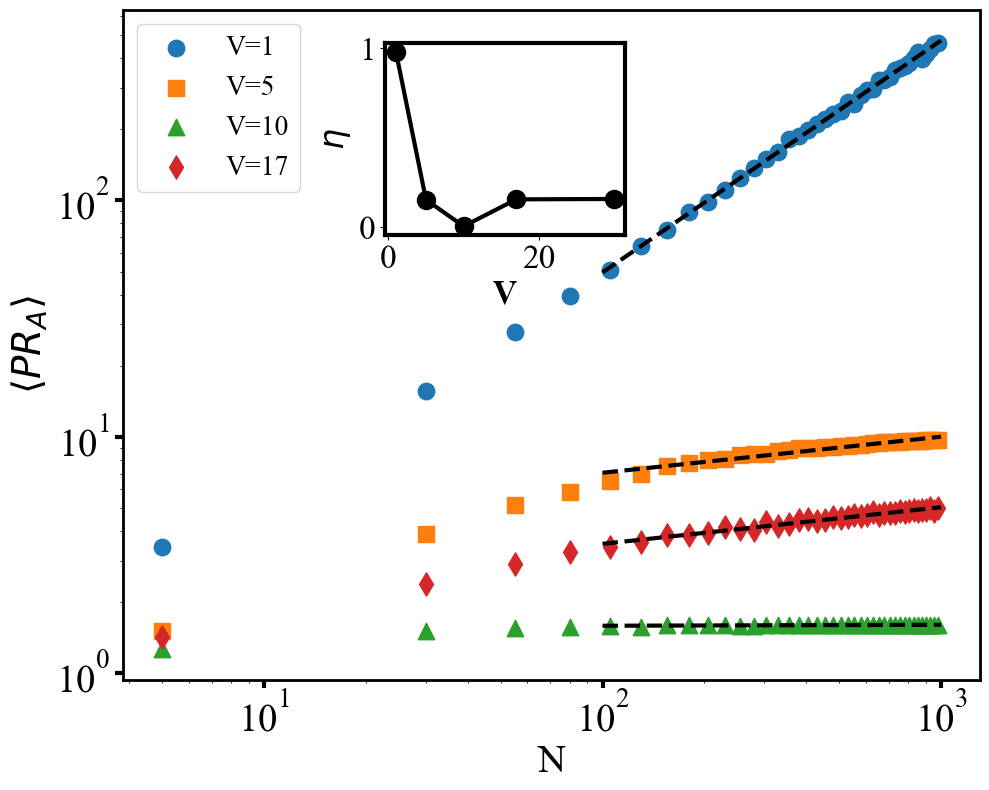}
    \caption{Log-log plots: $\langle PR_{A} \rangle$ vs. N for $V=1,5,10,17$ ($\epsilon=0.5$ and $t'=5$). The black dashed lines show the fitting to the functional form $\langle PR_{A} \rangle\sim N^{\eta}$.} Inset: Scaling exponent $\eta$ vs. V.
    \label{fig:eta_expnt}
\end{figure}

To analyze the localization character of the $\epsilon$-significant eigenkets, we now compute $IPR_A$ for these states. Fig. \ref{fig:IPR_ind_V} shows the dependence of $IPR_A$ on $V$ for the $\epsilon$-significant eigenkets with $\epsilon$ = 0.3, 0.5, and 0.7.
These calculations are performed for a total system size 2N = 1000, considering three different values of $t'$, namely $t'$ = 1, 5, and 10. The results clearly indicate that for small $t'$ (in particular, $t'=1$), most of the $\epsilon$-significant eigenkets are not localized within the subsystem $A$. As we increase $t'$ beyond a critical value $t'_c$, we see the appearance of a localized phase for intermediate values of $V$. 
In this regime,
almost all the $\epsilon$-significant eigenkets are localized up to a good extent but the states in the middle of the spectrum show stronger localization. 
 
For a given $t'>t'_c$, the subsystem shows localization when $V$ is above a lower bound (denoted by $V_1$) and below an upper bound (denoted by $V_2$). The lower and upper bounds for localization, i.e., $V_1$ and $V_2$, depend on the fixed value of $t'$. We calculate these two bounds in the following way.
For a state in the middle of the spectrum, we choose $V_1$ in such a way that $IPR_A<0.5$ for $V<V_1$ and $IPR_A\ge0.5$ for $V\ge V_1$. Similarly, for the same state, we choose $V_2$ in such a way that $IPR_A>0.5$ for $V<V_2$ and $IPR_A\le0.5$ for $V\ge V_2$. The plots for $V_1$ and $V_2$ as functions of $t'$ can be seen in Fig. \ref{fig:V1_V2}. These numerical results also help us find $t'_c$ by extrapolating $V_1$ and $V_2$ till they become equal. This extrapolation is performed by fitting the numerical data with a quadratic polynomial. From this procedure, we estimate $t'_c\approx4.4$. It is interesting to note that, as $t'$ increases beyond the critical value $t'_c$, the value of $V_2$ increases, while the value of $V_1$ decreases. The decrease in $V_1$ with increasing $t'$ can be understood from the fact that, in the limit of very large $t'$, the subsystem $A$ tends toward a fully localized phase, since the eigenstates are localized along the rungs of the ladder, for any finite $V$.

The analysis of the subsystem for $V>V_2$ (and $t'>t'_c$) reveals a new phase which we call a weakly delocalized phase with the fractal character. In this parameter regime, some states exhibit delocalization properties (having small $IPR_A$), while other states (near the edges) are found to be well localized (having large $IPR_A$). Such reentrance of the delocalized phase at large values of the quasiperiodic potential strength $V$ has recently been observed in several one-dimensional models \cite{tapan_reentrant1,tapan_reentrant2}.

For completeness, we also look into the energy-resolved IPR for the subsystem $A$, bath $B$ and the full system. The results may be found in Appendix \ref{ER_IPR}. In analyzing the localization properties of subsystem $A$, the role of the admixture of localized and delocalized (weakly delocalized) states is explained in the same appendix.

\subsection{$\langle PR_A\rangle$: Average Participation Ratio for subsystem}
To better understand the localization properties of different phases of the subsystem, we next perform the finite size scaling of the average participation ratio, $\langle PR_A\rangle$, where the average is taken only over the $\epsilon$-significant eigenkets:
\begin{equation} \label{pr_avr}
    \langle PR_A\rangle =\frac{1}{|\mathcal{S}_{\epsilon}|}\sum_{n \in \mathcal{S}_{\epsilon}}\frac{1}{IPR^{(n)}_A}.
\end{equation}
Here, $IPR^{(n)}_A$ is defined in Eq. \ref{IPR_eq} and the sum is over all the 
indices corresponding to the $\epsilon$-significant eigenkets in $\mathcal{S}_{\epsilon}$. We recall that $\mathcal{S}_{\epsilon}$ is the set of the $\epsilon$-significant eigenkets and $|\mathcal{S}_{\epsilon}|$ denotes the number of elements in the set (also see Eq. \ref{nu_eps}). For large subsystem size $N$, we expect the following finite size scaling law to hold:
\begin{equation} \label{fsc_eta}
   \langle PR_A \rangle \sim N^{\eta}. 
\end{equation}
This is a special case of a more general scaling relation which can be used for studying the multifractal nature of the subsystem: 
\begin{equation} \label{fractal_dim}
  \frac{1}{|\mathcal{S}_{\epsilon}|}\sum_{n \in \mathcal{S}_{\epsilon}}\left[\frac{1}{\sum_{j\in A}\lvert\braket{j|\psi_n}\rvert^{2q}}\right]\sim N^{D_q(q-1)}.
\end{equation}
Here, $D_q$ is the fractal dimension of the subsystem in a given phase. Our special case, as presented in Eq. \ref{fsc_eta}, corresponds to $q=2$ and $D_q=\eta$. The subsystem is considered to be in the delocalized (localized) phase when $\eta=1$ (0). The subsystem is said to be in a fractal phase if $0<\eta<1$. 
We point out here that the introduction of the fractal dimension $\eta$ (or $D_q$) for a subsystem closely parallels its definition for the full system, where it appears in the scaling relation of the average inverse participation ratio (IPR) of the entire system \cite{dynamics_2, tapan_reentrant2, Evers08}.

Fig. \ref{fig:PR_L} shows how $\langle PR_A \rangle$ scales with the subsystem size (N) for $V$ = 1, 10 and 17 (with fixed $t'$ = 5 and $\epsilon$ = 0.5). Corresponding log-log plots, as seen in Fig. \ref{fig:eta_expnt}, demonstrate that the quantity of interest obeys $\langle PR_A \rangle \sim N^{\eta}$ scaling in general. We emphasize that even when the subsystem hosts both delocalized (weakly delocalized) and localized states for certain values of the model parameters, this scaling relation remains valid. Further details are provided in Appendix~\ref{ER_IPR}.

We find that $\eta=1$ when $V=1$ (i.e., when $V<V_1$). This result confirms that the subsystem is in delocalized phase in the regime where $V$ is small. For $V=10$ (i.e., when $V_1<V<V_2$), we find $\eta \approx 0$. This shows that the subsystem $A$ is in a localized phase in the intermediate range of $V$. When $V=17$ (i.e., when $V>V_2$), we get $0<\eta<1$. This indicates that the corresponding phase is a fractal. Interestingly, when $V=5$ (i.e., when $V$ is close to $V_1$), we also get $0<\eta<1$. This observation suggests the presence of a narrow band along the transition line between the delocalized and localized phases, within which the subsystem exhibits a fractal phase.

\subsection{Subsystem properties for $t'<t'_c$}
So far we have mainly focused on the subsystem static localization properties for $t'>t'_c$. When $t'<t'_c$, the similar analysis of $IPR_A$ and $\langle PR_A \rangle$ shows that there is no localized phase and we have only two phases - a delocalized phase and a weakly delocalized (or fractal) phase. The relevant results may be found in Appendix \ref{appendix_t'=3} (especially, see Figs. \ref{fig:IPR_tp_3} and \ref{fig:PR_tp_3}).

\begin{figure}
    \centering
    \includegraphics[width=0.84\linewidth]{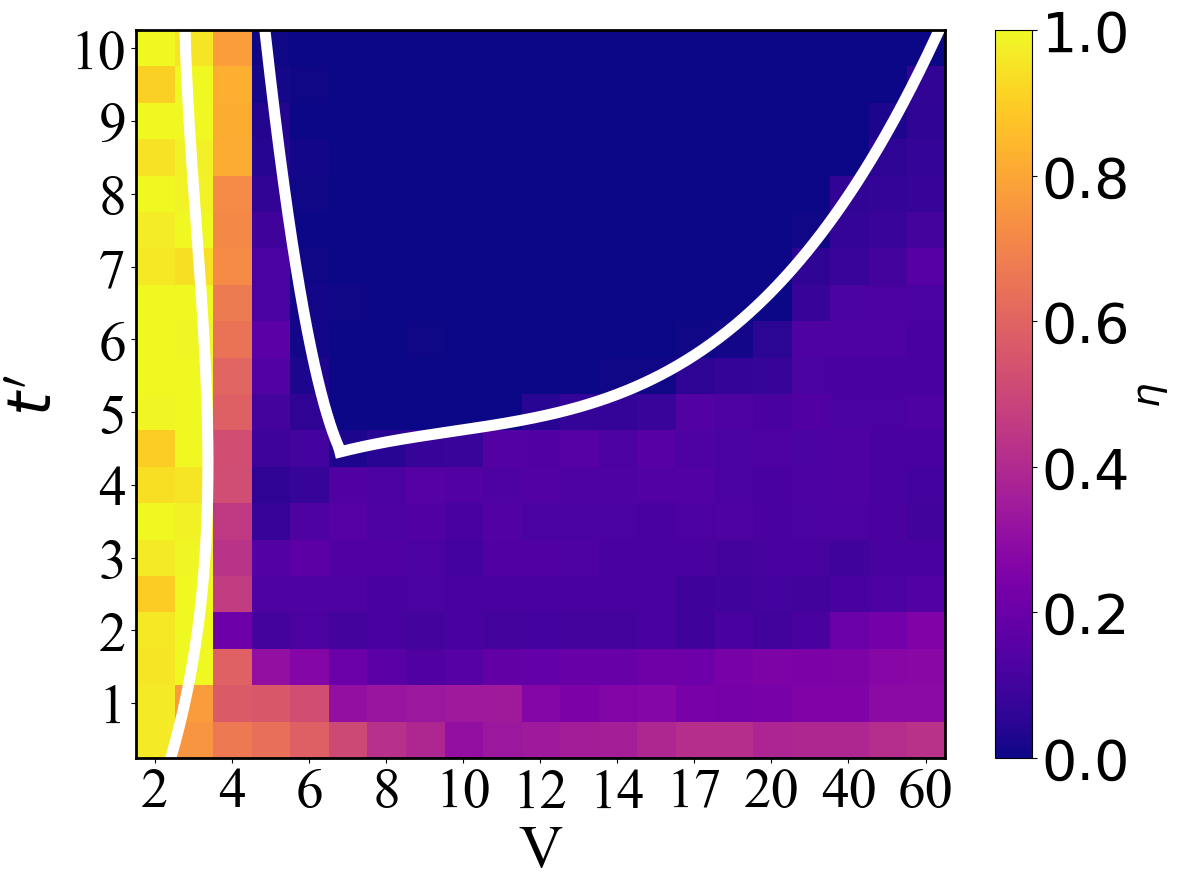}
    \caption{Contour plot of the scaling exponent $\eta$ as a function of $t'$ and V. Expected phase boundaries are indicated by white lines (see also Fig. \ref{fig:phases}).}
    \label{fig:contour_eta}
\end{figure}

\subsection{A contour plot for scaling exponent $\eta$}\label{contou_eta}
To gain a clearer understanding of the static phases of subsystem $A$, we evaluate the scaling exponent $\eta$ from the expected relation $\langle PR_{A} \rangle \sim N^{\eta}$, over a carefully chosen grid in the $t'-V$ plane. 
The grid is defined as follows: for $t'$, we consider values ranging from $0.5$ to $10.0$ in steps of $0.5$ (20 values in total); for $V$, we take the set of 21 values consisting of $2.0$ to $15.0$ in steps of $1.0$, $17.0$, $19.0$, and $20.0$ to $60.0$ in steps of $10.0$. To extract the value of $\eta$, we calculate $\langle PR_{A} \rangle$ for $N =$ 300, 400, 600, 800 and 1000. The values of $\eta$ at different grid points are presented as a contour plot in Fig. \ref{fig:contour_eta}.

This contour plot clearly shows what we expected earlier from our detailed calculations on some carefully chosen points in the parameter plane. For $t'>t'_c$, we have a localized phase ($\eta\sim 0$) in the intermediate values of the AA potential strength $V_1<V<V_2$. For $V<V_1$ and $V>V_2$, we respectively get a delocalized phase ($\eta\sim1$) and a weakly delocalized fractal phase ($0<\eta<1$). Interestingly, we also get a narrow fractal phase along the line separating delocalized and localized phases. 
When $t' < t'_c$, no evidence of a localized phase is observed. For small $V$, the system remains in a delocalized phase, while for intermediate to large $V$, the subsystem enters a weakly delocalized fractal phase. Within this fractal regime, for a fixed $V$, the scaling exponent $\eta$ increases as $t'$ decreases, and $\eta \to 1$ as $t' \to 0$. We recall that the subsystem $A$ stays in a delocalized phase ($\eta\sim1$) on the $t'=0$ axis.

\subsection{Understanding reentrance phenomenon \label{sec:reentrance}}
For $t' > t'_c$, the subsystem $A$ exhibits three major distinct phases as $V$ increases: a delocalized phase, followed by a localized phase, and finally a weakly delocalized (fractal) phase. This sequence represents a form of reentrance behavior, which can be understood intuitively as follows.

For small $V<2$ with $t'=0$, both subsystem $A$ and bath $B$ are delocalized (eigenkets spread within $A$ or $B$). Introducing a finite coupling $t'>0$ keeps both the subsystem and bath delocalized, though now the eigenkets extend across the full system $A+B$ (see Appendix~\ref{ER_IPR}).

For $V>2$ with $t'=0$, bath $B$ becomes localized while subsystem $A$ retains $N$ delocalized states. Increasing $t'$ beyond a certain value then hybridizes these states in $A$ into rung-localized dimers, whose projected parts are localized in $A$ as reflected by large $IPR_A$.

At larger $V$ (keeping $t'$ fixed), bath $B$ develops strongly localized states, while the remaining eigenstates in $A$ weakly couple to $B$. The projected such states form the weakly delocalized (fractal) phase.

\begin{figure}
    \centering
    \includegraphics[width=0.84\linewidth]{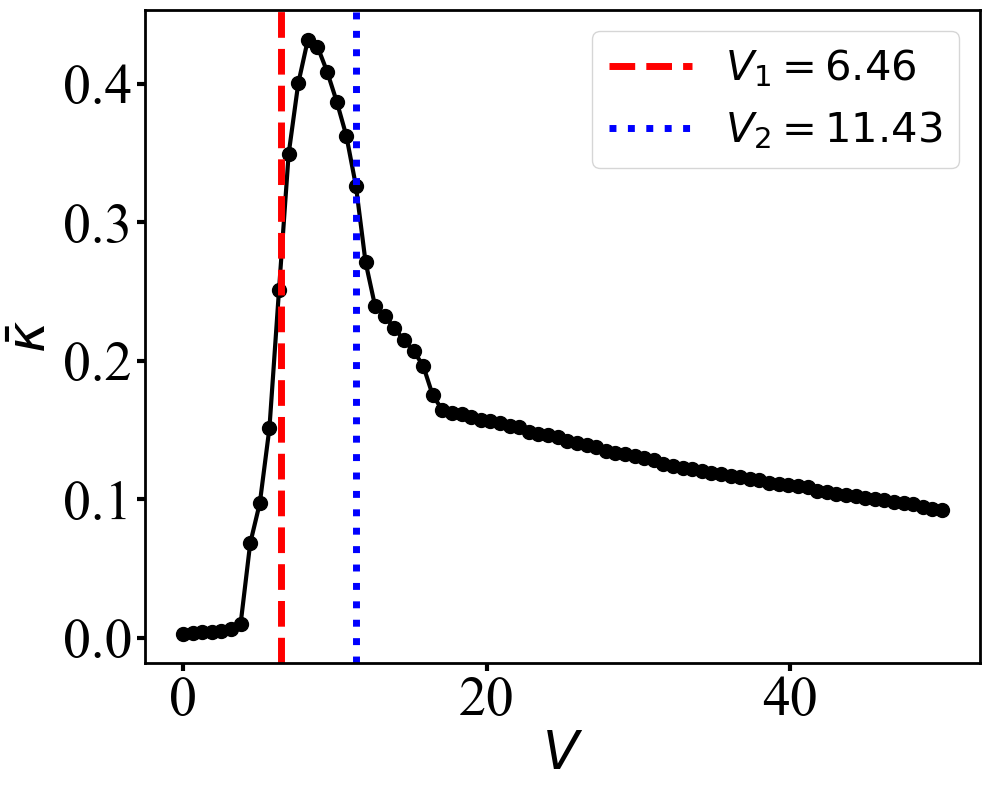}
    \caption{Plot of $\overline{\kappa}$ as a function of V with $t'=5$. For the calculations, we consider the subsystem size $N=500$ and take only $\epsilon=0.5$ significant states.}
    \label{fig_kappa}
\end{figure}

\begin{figure*}[t]
\includegraphics[width=0.88\textwidth]{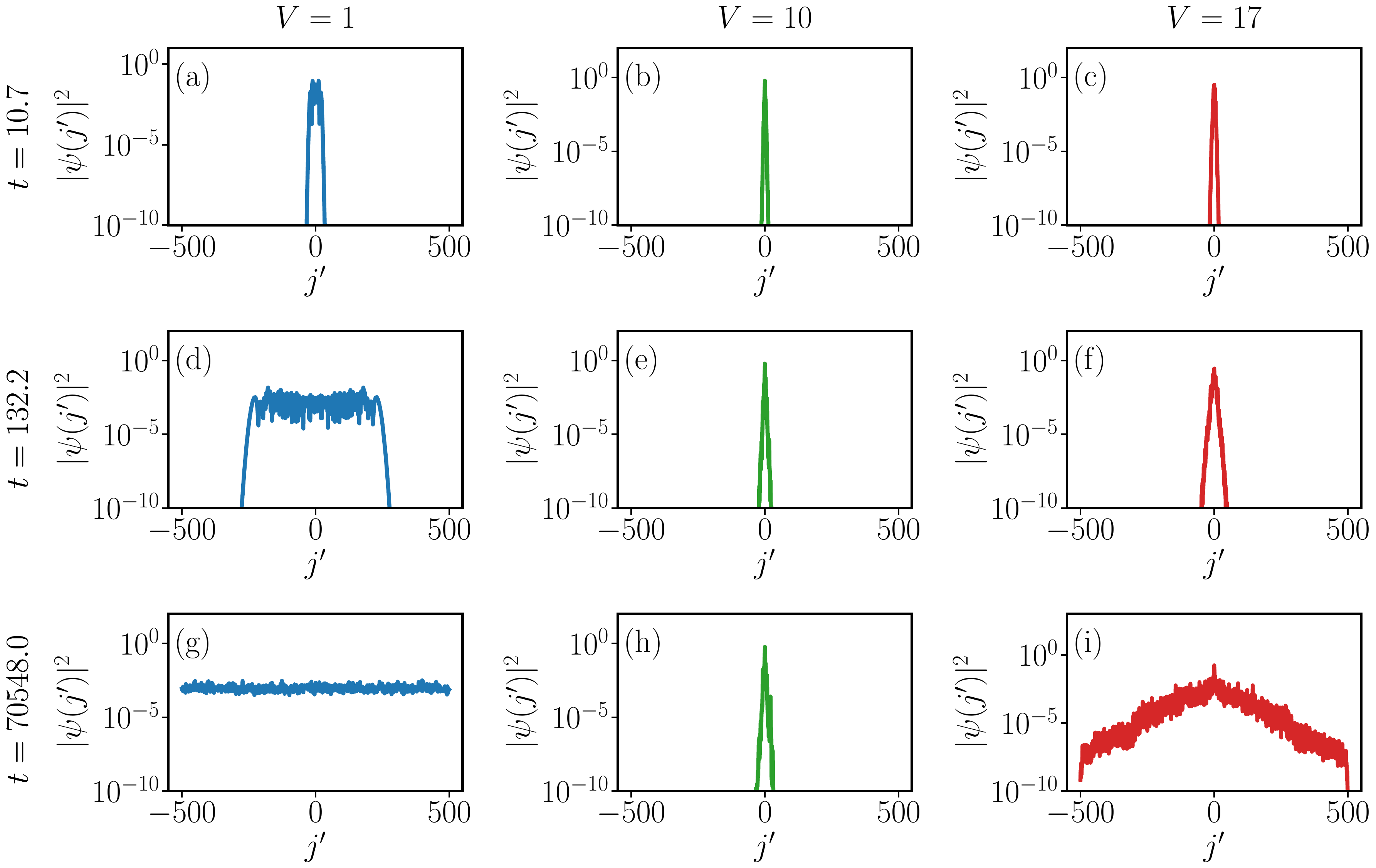}
\centering
\caption{Probabilty distribution $|\psi(j')|^2$ vs. $j'$ plot for different $V$ and different instant of time $t$, for  $N=1000$. The particle is initiated from the middle of the subsystem $A$, i.e., $j=N/2$, and here $j'=j-N/2$.}
\label{Fig4}
 \end{figure*}

To investigate the formation of dimer states — which are responsible for the intermediate localized phase — we introduce a measure that quantifies bond localization. For an eigenket $\ket{\xi}$, the bond localization parameter $\kappa_{\xi}$ is define as
\begin{equation} \label{kappa_para}
    \kappa_{\xi} = \sum_{i=1}^N|\bra{\xi}c_i^\dagger c_{N+i} + c_{N+i}^\dagger c_{i}\ket{\xi}|^2.
\end{equation}
We recall that the $i$th site of subsystem $A$ is coupled to the $(N+i)$th site of bath $B$ through the hopping parameter $t'$. As shown in Appendix \ref{bond_local}, the bond localization parameter takes the value $\kappa_{\xi} = 1$ for an ideal dimer state, such as
$\ket{\xi} = \frac{1}{\sqrt{2}}\big(\ket{i} \pm \ket{N+i}\big)$.
In the thermodynamic limit, $\kappa_{\xi} < 1$ for any non-ideal dimer state, while $\kappa_{\xi}$ vanishes for states that are fully delocalized within $A$, within $B$, or across both subsystems. Between two states, the state with the larger value of $\kappa$ is expected to have more characteristics of a dimer state.

To gain insight into how this bond localization parameter changes as a function of $V$ for a given $t'$, we plot average of this parameter, defined as follows:
\begin{equation} \label{avkappa_para}
    \overline{\kappa} = \frac{1}{|\mathcal{S}_{\epsilon}|}\sum_{\ket{\xi}\in \mathcal{S}_{\epsilon}}\kappa_{\xi},
\end{equation}
where $\mathcal{S}_{\epsilon}$ is the set of $\epsilon$-significant states as defined earlier (see Sec. \ref{sec3a}). The plot of $\overline{\kappa}$ as a function of $V$ (with $t'=5>t'_c$) can be found in Fig. \ref{fig_kappa}. This result clearly demonstrates that there is indeed localization along the bonds (rungs) in the intermediate region of $V$ where we observe the subsystem localization in the reentrance phenomenon.

\section{Dynamic properties}\label{sec4}

\begin{figure}[t]
    \centering
\includegraphics[width=0.44\textwidth]{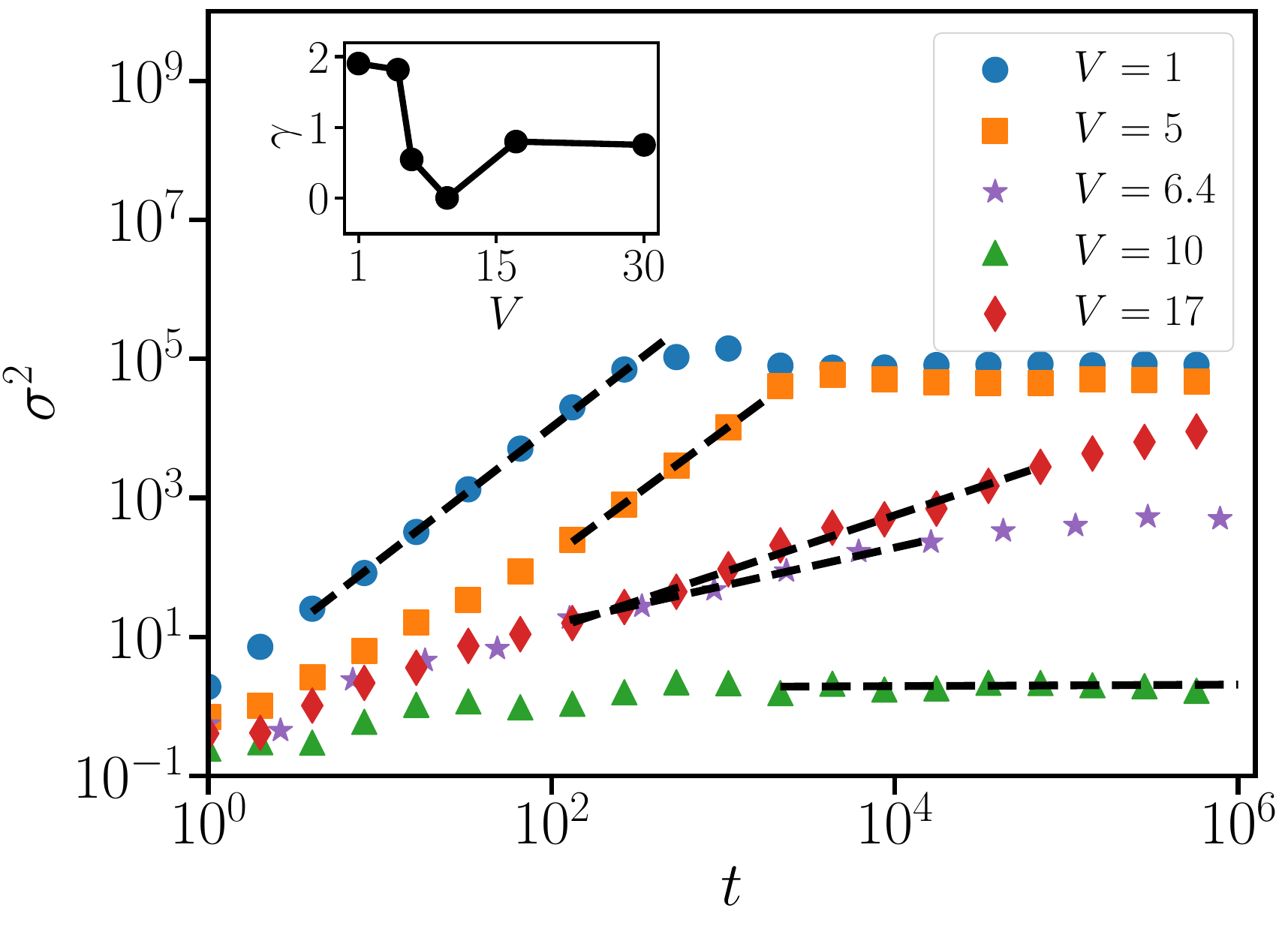}
\caption{$\sigma^2$ vs $t$ plot for different $V$ for  subsystem size $N=1000$ and $t'=5 >t'_c$. The black dashed lines show the fitting to the functional form $\sigma^2\sim t^\gamma$. The inset shows the variation of the exponent $\gamma$ on $V$.}
\label{Fig5}
\end{figure}
\begin{figure}[t]
    \centering
\includegraphics[width=0.44\textwidth]{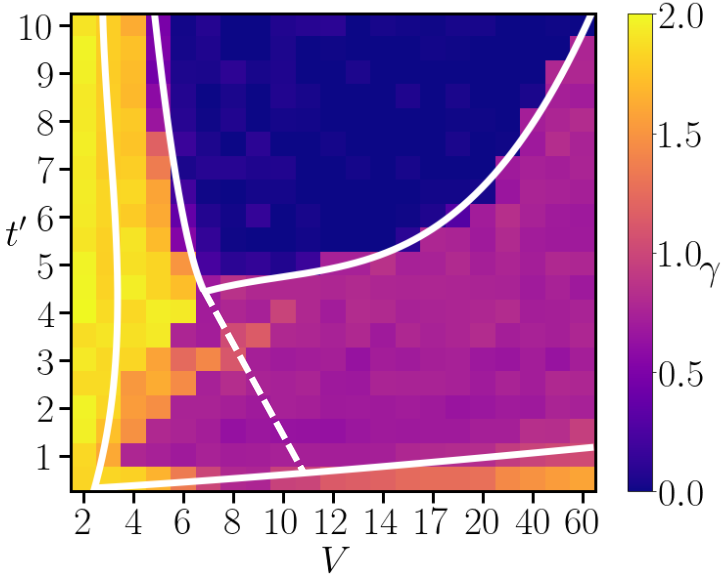}
\caption{Contour plot of the scaling exponent $\gamma$ as a function of $t'$ and $V$. Expected phase boundaries are indicated by white lines (see also Fig. \ref{fig:phases}). }
\label{Fig5_1}
\end{figure}
\begin{figure}[t]
    \centering
\includegraphics[width=0.43\textwidth]{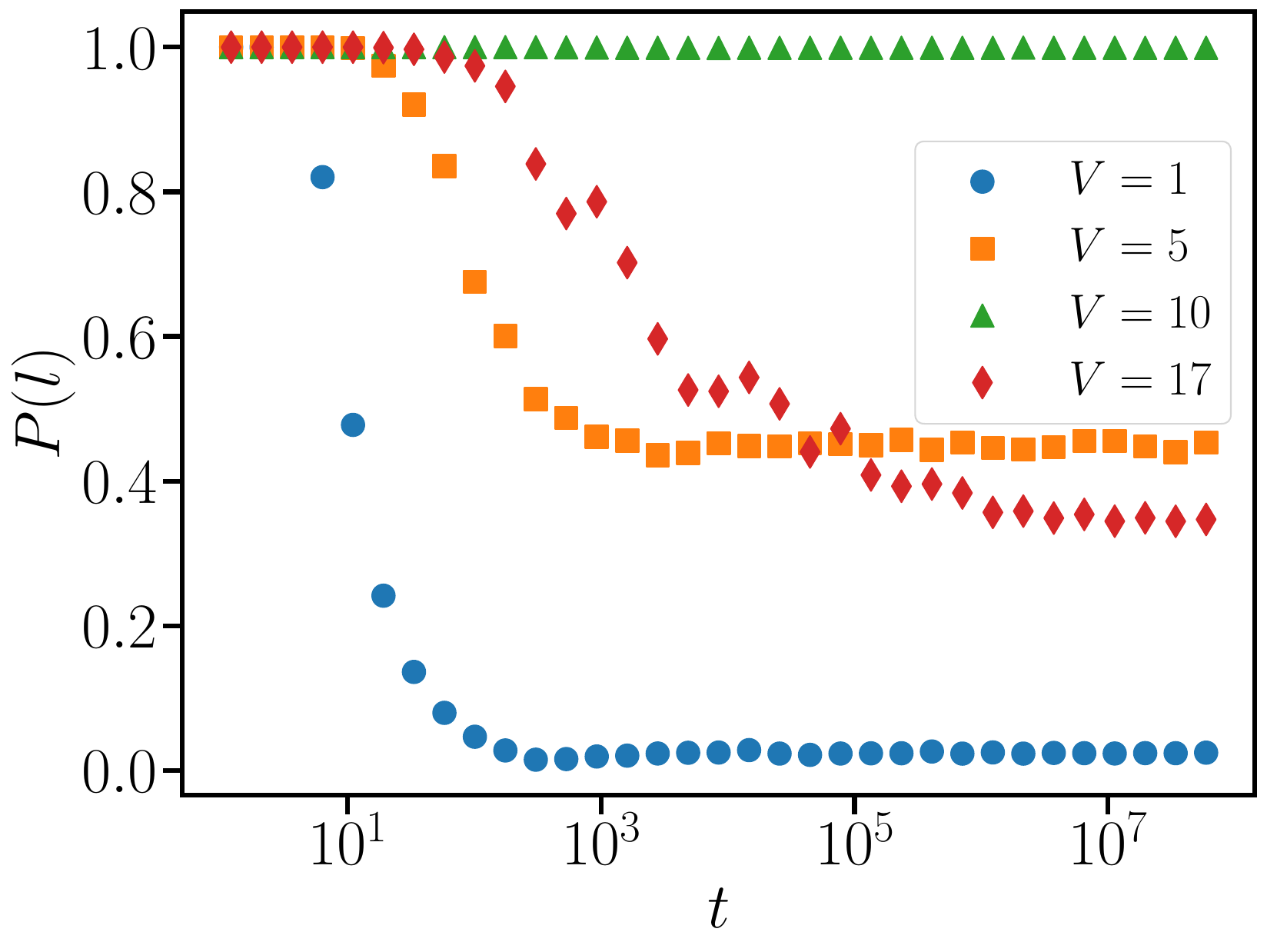}
\caption{Variation of $P(l)$, the Probability of finding the particle in the region $N/2-l$ to $N/2+l$, with time  $t$ for different $V$. Results are for $N=1000$,  $l=10$, and $t'=5$. Plots are generated using local time averaging to smooth out the data.}
    \label{Fig6}
\end{figure}

In this section, we study the projected wave packet dynamics in subsystem $A$. We initialize the system by placing one particle at the center of subsystem $A$, and then let it evolve under the total Hamiltonian $H$ as 
\[
|\xi(t)\rangle = e^{-iHt}|\phi_0\rangle,
\]
where the initial state is given by $|\phi_0\rangle = |1_{N/2}\rangle$.

We focus on the normalized time-evolved state projected onto subsystem $A$, defined as
\[
|\psi(t)\rangle = \frac{\hat{P}_A|\xi(t)\rangle}{\|\hat{P}_A|\xi(t)\rangle\|},
\]
which is the same as Eq.~\eqref{ket_A}, and study its dynamics.

Here, we first focus on $t'=5>t_c'$.
Figure~\ref{Fig4} shows the time evolution of the probability distribution $|\psi(j,t)|^2 = |\langle j|\psi(t)\rangle|^2$ for $V = 1$, $10$, and $17$. According to our static results, these three values of $V$ correspond to completely delocalized, completely localized, and weakly delocalized (fractal) phases, respectively. We observe that for $V = 1$, $|\psi(j,t)|^2$ spreads significantly faster across subsystem $A$ compared to the case of $V = 17$. In contrast, for $V = 10$, the probability distribution remains nearly unchanged over time, indicating a fully localized phase. Next, we quantify the spread of the probability distribution using different measures.

\subsection{MSD of the projected time-evolved state}\label{variance_vs_time}

First, we quantify the spread of the projected time-evolved state in subsystem $|\psi(j,t)|^2$ using the variance $\sigma^2$ of the probability distribution. 
The variance $\sigma^2$ of this probability distribution is defined as,
\begin{equation}
    \sigma^2(t)= \sum_{j=1}^N j^2 |\psi(j,t)|^2  - \left[\sum_{j=1}^{N} j |\psi(j,t)|^2 \right]^2.
\end{equation}
Typically, the long-time growth of $\sigma^2$ scales as
\begin{equation}
\sigma^2(t) \sim t^{\gamma},
\end{equation}
with $0 \leq \gamma \leq 2$. Ballistic transport corresponds to $\gamma = 2$, while $1 < \gamma < 2$ ($0 < \gamma < 1$) indicates superdiffusive (subdiffusive) behavior. The case $\gamma = 1$ represents diffusive transport, and in the localized phase one finds $\gamma = 0$.

In Fig.~\ref{Fig5}, we plot $\sigma^2$ as a function of time $t$ for different values of $V$ (with $t' = 5$ fixed). In the completely localized phase (e.g., $V = 10$), we observe that $\sigma^2$ remains constant over time, as expected. In contrast, in other regimes, $\sigma^2$ grows as $\sigma^2 \sim t^\gamma$ until it saturates due to finite-size effects.
For small values of $V$ (e.g., $V = 1$), we find the exponent $\gamma = 2$, indicating ballistic transport. As $V$ increases, we enter into a super-diffusive regime where $1 < \gamma < 2$ (see results for $V = 5$), followed by a sub-diffusive regime with $\gamma < 1$ (see results for $V = 6.4$), and a completely localized phase. 
Upon further increasing $V$ (e.g., $V = 17$), we enter a weakly delocalized (fractal) regime, as identified in our static calculations, where the dynamics becomes sub-diffusive, i.e., $\gamma < 1$. 
The variation of the exponent $\gamma$ with $V$ is shown in the inset of Fig.~\ref{Fig5}. We discuss the robustness of the exponent $\gamma$ by varying the system size in Appendix~\ref {crossover regime}. An alternative validation of subdiffusive transport, based on the scaling of the projected probability distribution, is presented in Appendix~\ref{appendix_psi_scaling}. 

\subsection{A contour plot for scaling exponent $\gamma$}\label{contour plot gamma}
To gain a clearer understanding of the dynamic phases of subsystem $A$, we evaluate the scaling exponent $\gamma$ from the expected relation $\sigma^2 \sim t^{\gamma}$, over a carefully chosen grid (same as mentioned in Sec .~\ref {contou_eta}) in the $t'-V$ plane. To obtain the exponent $\gamma$, we have fitted the long-time regime (before saturation) data of $\sigma^2$ vs $t$ plot for $N=1000$. The value of $\gamma$ at different grid points is presented as a contour plot in Fig.~\ref{Fig5_1}. 

As indicated by our static calculations, for a fixed $t' > t'_c \approx 4.4$, we find that the exponent $\gamma \approx 0$ within a finite interval between $V_1$ and $V_2$, signaling the presence of a localized phase, which is identified as $P2$ phase in Fig.~\ref{fig:phases}. We further note that $V_1$ decreases slightly while $V_2$ increases with increasing $t'$. This trend suggests that in the limit $t' \to \infty$, the subsystem $A$ becomes completely localized for any finite value of $V$, which is consistent with the intuitive picture of eigenstate dimerization in the total Hamiltonian $H$. On the other hand,  the exponent $\gamma \approx 2$ for $V \ll V_1$, indicating a ballistic phase (which is identified as a part of  $P1$ phase in Fig.~\ref{fig:phases}), and $\gamma < 1$ for $V > V_2$, suggesting a sub-diffusive phase (which is identified as a part of  $P3$ phase in Fig.~\ref{fig:phases}).

If $t'<t_c'$ but sufficiently large, we find a ballistic and the sub-diffusive phase in the small and the large $V$ regime, respectively (see Appendix.~\ref{appendix_t'=3} for details), which is once again identified as a part of  $P1$ phase and $P3$ phase in Fig.~\ref{fig:phases}.  If $t'/V\ll 1$, we observe that even at large $V$ the system can exhibit a super-diffusive phase with $\gamma > 1$ (which is identified as $P5$ phase in Fig.~\ref{fig:phases}). Moreover, for extremely small $t'/V$, i.e.,  $t'/V\to 0$, we find $\gamma \to 2$, indicating even the existence of a ballistic phase. This ballistic phase is not visible on the scale of Fig.~\ref{Fig5_1}; therefore, in Appendix~\ref{very small t'}, we present a zoomed-in version of the small-$t'$ regime of the phase diagram to clarify this feature. Note that in the strict $t'=0$ limit, the dynamics are trivially ballistic, as it is entirely governed by the Hamiltonian $H_A$ (see Eq. \ref{ham_eqn}).

We also identify an intermediate crossover regime (which is identified as $P4$ phase in Fig.~\ref{fig:phases}) between the ballistic and localized phase (for $t' > t'_c$), as well as between the ballistic and sub-diffusive phase (for $t' < t'_c$). 
In this phase, the exponent $\gamma$ exhibits a strong dependence on both the system size and the model parameters. Our $N=1000$ data suggests the existence of both super-diffusive and sub-diffusive phases within this crossover regime. However, a clear phase boundary between these phases is extremely difficult to establish using finite-size numerics. Further details are provided in Appendix~\ref{crossover regime}.

\begin{figure}[t]
    \centering
\includegraphics[width=0.44\textwidth]{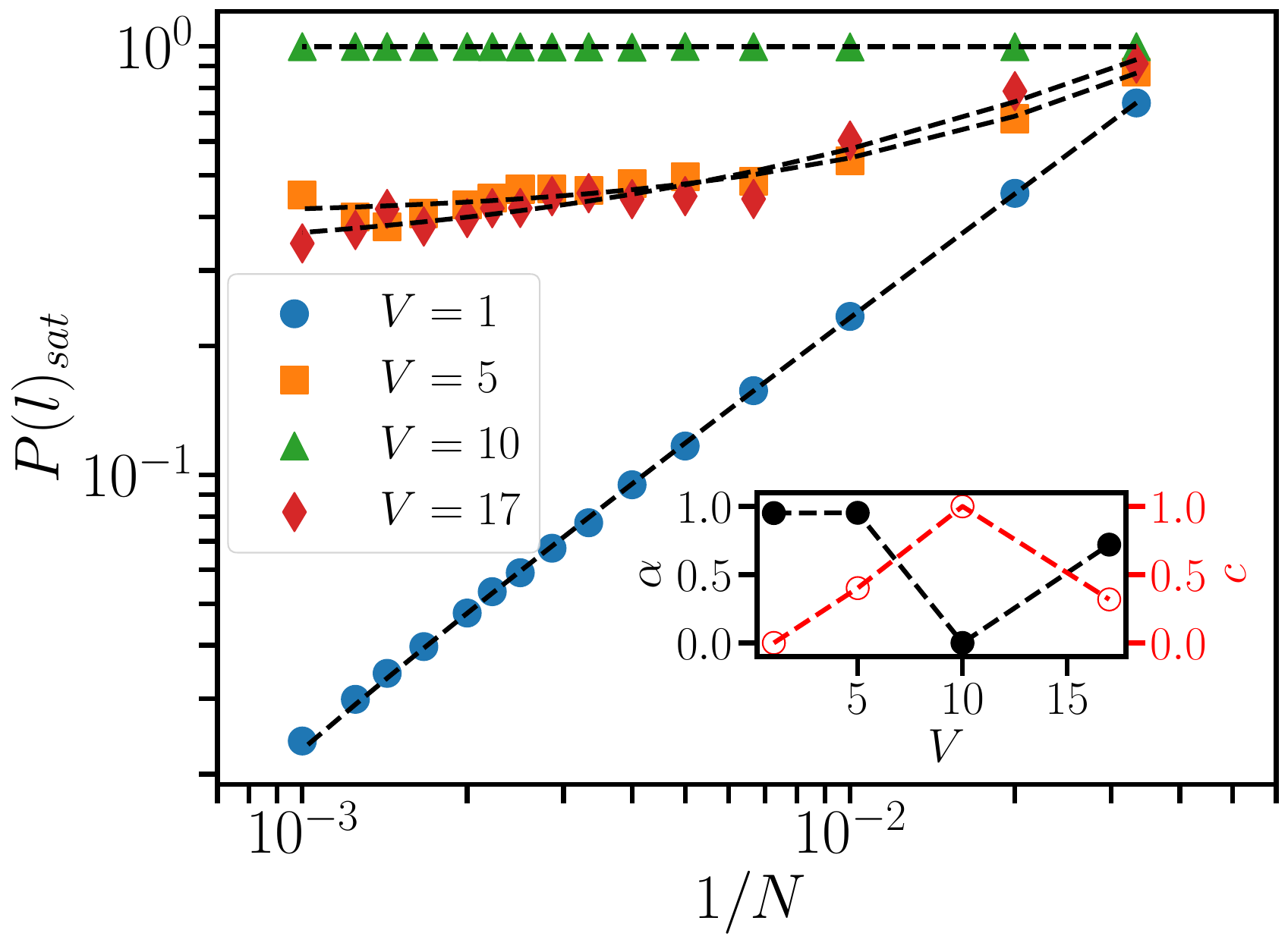}
\caption{Scaling of $P(l)_{sat}$, the long time average of $P(l)$, with inverse of the subsystem size $1/N$. The black dashed line represents the fitting of the data with the functional form $P(l)_{sat} \sim (1/N)^{\alpha} + c$. The inset shows the variation of $\alpha$ and $c$ with $V$. Results are for $2l=20$, and $t'=5$.}
\label{sat_scale}
\end{figure}

\subsection{Memory of initial state}\label{memory measure}

Further, we investigate another measure, $P(l,t)$, defined as
\begin{equation}
    P(l,t)= \sum_{j=N/2-l}^{N/2+l} |\psi(j,t)|^2,
\end{equation}
where $l \ll N$. This quantity represents the total probability of finding the particle at time $t$ within a neighborhood of width $2l$ centered around its initial position. For our choice of initial state, we have $P(l,0) = 1$. This measure provides insight into how much memory of the initial state is retained during the dynamics. 
In the completely localized phase, one expects $P(l,t) \approx 1$ for all times, provided $l$ is larger than the typical localization length. On the other hand, in a completely delocalized phase, one would expect that at sufficiently large times $|\psi(j,t)|^2 \sim \frac{1}{N}$ for all $j$, implying that $P(l,t)$ saturates to $P(l) \approx \frac{2l}{N}$. This scales to zero in the thermodynamic limit ($N \to \infty$), indicating a complete loss of memory of the initial state.
Figure~\ref{Fig6} confirms this behavior for $t'=5>t'_c$: for $V = 1$ (completely delocalized) and $V = 10$ (completely localized), $P(l = 10, t)$ behaves as expected—decreasing rapidly with time and saturating to a small value for $V = 1$, while remaining close to $1$ for all times in the case of $V = 10$. Interestingly, for  $V = 5$ and $V = 17$, $P(l,t)$ decreases at early times but saturates to a significantly higher value than in the delocalized case ($V = 1$), indicating partial memory retention.

To investigate  further whether the initial memory persists in the thermodynamic limit, we perform a finite-size scaling analysis of the long-time averaged value of $P(l,t)$, defined as
\begin{equation}
    P(l)_{\text{sat}} = \lim_{T \to \infty} \frac{1}{T} \int_{0}^{T} P(l,t)\, dt.
\end{equation}
Figure~\ref{sat_scale} shows the scaling of $P(l)_{\text{sat}}$ with subsystem size $N$ for different values of $V$ and fixed $t'=5$. We fit the data using the function $\text{constant}/N^{\alpha} + c$, and the variation of the fitting parameters $\alpha$ and $c$ with $V$ is shown in the inset of Fig.~\ref{sat_scale}.
In the completely delocalized phase ($V = 1$), we find $\alpha \approx 1$ and $c \approx 0$, as expected. In the completely localized phase, $P(l)_{\text{sat}}$ remains nearly constant with $N$. Interestingly, for $V = 5$ (super-diffusive) and $V = 17$ (sub-diffusive), the scaling behavior is similar, but the fitted constant $c \neq 0$ in both cases. This indicates that a finite portion of the initial state's memory survives even in the thermodynamic limit.\\  

\section{Analytical understanding of results}\label{sec:analytic}
 
 \subsection{Toy model}
 \begin{figure}[t]
  \centering
\includegraphics[width=0.44\textwidth]{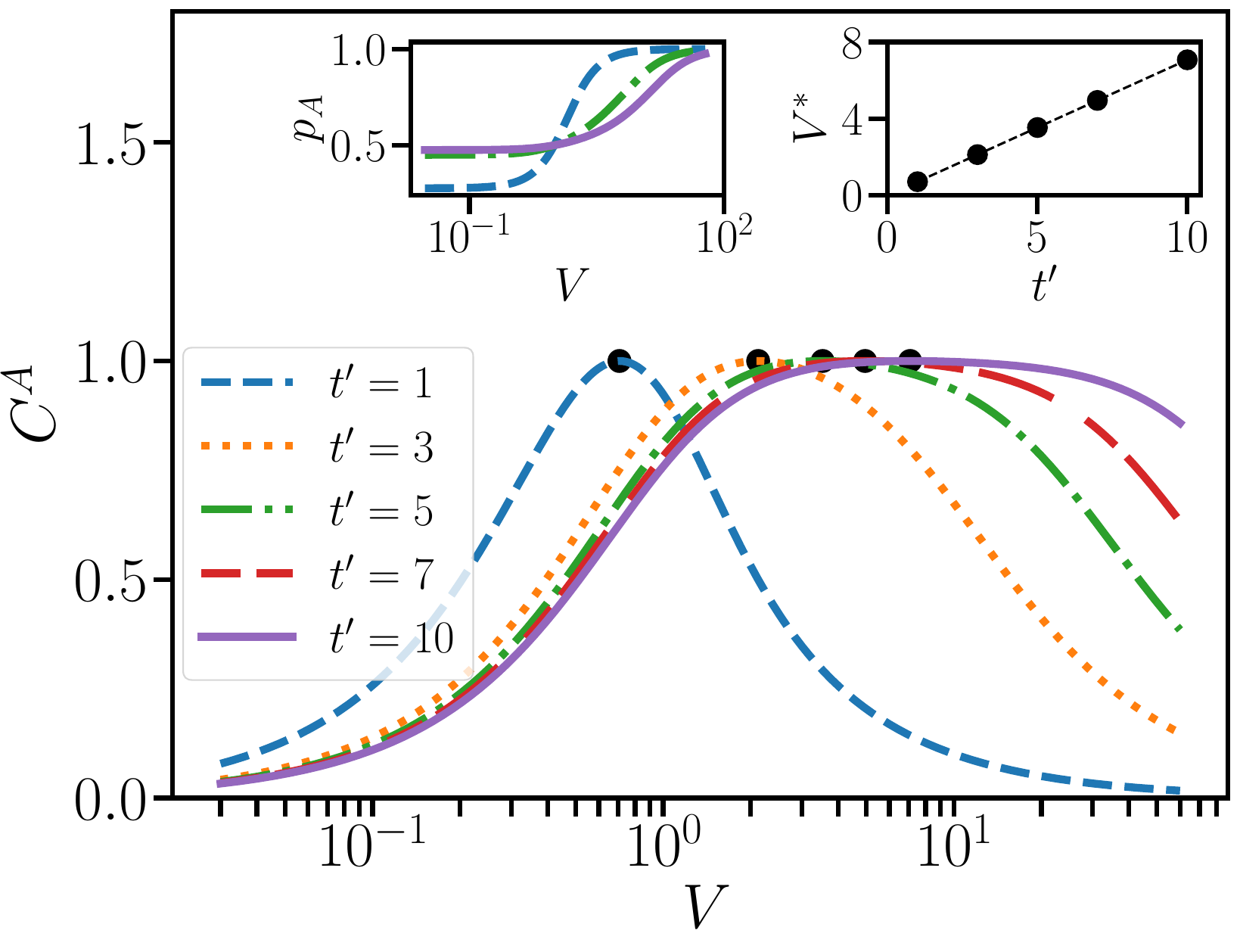}
\caption{A localization measure $C^A$ vs. $V$ plots for the most significant state of $H_{T}$ for different $t'$. The top-left inset shows the probability of a particle being at subsystem $p_A$ vs. $V$ for the same 
 state. The top-right inset shows the scaling of  $V^*$ (the $V$ for which $C^A$ attains a peak) with $t'$. }
\label{new_Fig9}
\end{figure}

Here, we propose a minimal toy model of four sites to qualitatively understand some of the numerical results obtained for the Hamiltonian in Eq.~\eqref{ham_eqn}. 
Primarily, we make three observations based on our numerical results when $t'>t'_c$:
(1) the subsystem becomes localized over an intermediate range of $V$,
(2) the extent of this localization regime ($V_1<V<V_2$) increases with $t'$, and perhaps most surprisingly (3) we observe a reentrance of the delocalized phase in the subsystem at larger $V$ (as discussed in Sec. \ref{sec:reentrance}).

One of our goals here is to investigate whether similar phenomena can also be observed in a minimal, fully analytically solvable toy model consisting of only four sites. In this setup, sites $1$ and $2$ constitute the subsystem $A$, while sites $3$ and $4$ form the bath $B$ (see Fig. \ref{fig:system} with $N=2$). Localization in $B$ is induced by assigning on-site potentials $+V$ and $-V$ to sites $3$ and $4$, respectively. To ensure complete localization in $B$ for any finite $V$, we set the hopping amplitude between sites $3$ and $4$ to zero. In the absence of coupling between $A$ and $B$ (i.e., for $t'=0$), this guarantees full localization within $B$.

The minimal toy model is described by the Hamiltonian, 
\begin{equation}
 H_T = -(c^{\dagger}_1c_2 + t' c_1^{\dagger} c_3 + t' c_2^{\dagger} c_4 + hc) + V(c^{\dagger}_3 c_3 - c^{\dagger}_4 c_4).\\ 
 \label{toymodel}
\end{equation}
If $t'=0$, the A and B parts of the system are completely decoupled. The two eigenstates in B are fully localized, and the two in A are maximally delocalized.
For those maximally delocalized eigenstates that lie in A, 
the probability of finding the particle on sites $1$ and $2$ remains equal.  
Next, we introduce a finite $t'$ and investigate how this scenario evolves with varying $V$. To quantify the degree of localization of an eigenstate within subsystem A, we define the following measure for an eigenstate $|\xi\rangle$:
\begin{equation}
    C^A = \frac{\big||\langle 1|\xi\rangle|^2 - |\langle 2|\xi\rangle|^2\big|}{|\langle 1|\xi\rangle|^2 + |\langle 2|\xi\rangle|^2}.
\end{equation}
Now, among the two sites in A, if $|\xi\rangle$ is completely localized on only one of the sites (either at site $1$ or $2$), then 
 $C^A=1$, and on the other hand, if it is completely delocalized (probabilities to be at site $1$ and $2$ are equal), then  $C^A=0$. 
Given $H_T$ is a $4\times 4$ matrix,  $C^A$ can be obtained analytically, and for the most significant state, 
\begin{equation}
    C^A=\frac{\big|(y-V)^2 -(y^2 -Vy-t'^2)^2\big|}{(y-V)^2 +(y^2 -Vy-t'^2)^2},
\end{equation}
where, $x=\sqrt{\frac{V^{2}}{2} + t'^{2} + \frac{\sqrt{V^{4} + 4 V^{2} t'^{2} - 2 V^{2} + 4 t'^{2} + 1}}{2} + \frac{1}{2}}$, and $y=\sqrt{\frac{V^{2}}{2} + t'^{2} - \frac{\sqrt{V^{4} + 4 V^{2} t'^{2} - 2 V^{2} + 4 t'^{2} + 1}}{2} + \frac{1}{2}}$. 

Figure~\ref{new_Fig9} (main panel) shows the variation of $C^A$ with $V$ for different values of $t'$. As $V$ increases, $C^A$ initially increases, reaches a peak at $V = V^*$. 
We obtain the analytical expression $V^* = \frac{t'}{\sqrt{2}}$ by solving the equation $\frac{d C^A}{dV}\big{|}_{V^*} = 0$. It is also straightforward to check that $C^A \to 1$ as $V \to V^{*}$, indicating a completely localized phase in subsystem A. For $V>V^{*}$, however, $C^A$ begins to decrease. This behavior closely mirrors the emergence of a localized phase at intermediate $V$ and the subsequent reentrance of the delocalized phase in the large-$V$ regime, as observed in the original Hamiltonian $H$. Moreover, we find that the decay of $C^A$ for $V>V^*$ becomes progressively slower with increasing $t'$, suggesting that the localized phase persists over a broader range of $V$ as $t'$ increases. This trend is consistent with our earlier observations for the ladder Hamiltonian.

The inset of Fig.~\ref{new_Fig9} shows the variation of the total probability that the particle remains in subsystem A for the eigenstate $|\xi\rangle$, given by
\begin{equation}
   p_A = |\langle 1|\xi\rangle|^2 + |\langle 2|\xi\rangle|^2,
\end{equation}
as a function of $V$. We find that as $V$ increases, this probability approaches $1$, indicating that the state becomes increasingly significant within subsystem A. 

Moreover, we have also studied a bond-localization measure, defined as  $\Omega_{\xi}=4(|\langle 1|\xi\rangle|^2  |\langle 3|\xi\rangle|^2+|\langle 2|\xi\rangle|^2  |\langle 4|\xi\rangle|^2$),  
for our toy model. This measure approaches $1$ ($0.5$) when the eigenstate is completely bond localized (delocalized). For a fixed $t'$, we find that $\Omega_{\xi}$ approaches $1$ within an intermediate range of $V$, signaling the onset of bond localization. However, upon further increasing $V$, $\Omega_{\xi}$ decreases again. This observation reinforces the connection between re-entrant behavior and bond localization, even within our minimal toy model.

\subsection{Effective model for the subsystem}
Here, we show that a model closely related to ours (Eq. \ref{ham_eqn}) can be reduced to a simple effective one-dimensional model for the subsystem. The model we consider here does not have the hopping term along the $B$ leg (the bath). The Hamiltonian of this model is the same as the one appears in Eq. \ref{ham_eqn} with $t_B=0$. The two models are expected to give similar results in the large $t'$ and $V$ limits. 

Consider that \(\ket{\psi}= \sum_{i\in A} a_i \ket{i} +\sum_{N+i\in B} b_{i} \ket{N+i} \) is an eigenstate of the Hamiltonian of the current model (where $t_B=0$). As per the current notation, the coefficient $a_i$ is associated with the site basis $\ket{i}$ of the subsystem $A$, and the coefficient $b_i$ is associated with the site basis $\ket{N+i}$ of the subsystem $B$. 
We recall that, the $i$th site in $A$ is directly connected to the $(N+i)$th site in $B$ by the hopping parameter $t'$ (see Hamiltonian in Eq. \ref{ham_eqn}). If $E$ is the eigevalue associated with $\ket{\psi}$, then 

\begin{align}
        -t'a_{i}+V_i b_i =& E b_i\label{eigB} \\
        -t_A a_{i+1}-t_A a_{i-1} -t' b_i=& E a_i\label{eigA}
\end{align}
Here $V_i=V\cos{(2\pi\beta i+\phi)}$, the AA potential as appears in Eq. \ref{ham_eqn}. From Eq. \ref{eigB}, we get:
\[b_i =\frac{-t' a_i}{(E-V_i)}.\]
We use this expression to replace $b_i$ in Eq. \ref{eigA}; this yields:
\begin{equation}
     -t_A a_{i+1}-t_A a_{i-1} +\frac{(t')^2 a_i}{(E-V_i)} =E a_i.
\end{equation}
One can recast this equation in the following form:
\begin{equation}
     -t_A a_{i+1}-t_A a_{i-1} + V_{i}^{eff} a_i =E^{eff} a_i.
\end{equation}
Here, $E^{eff}=E-1/\alpha$ and
\[V^{eff}_i= \frac{\lambda \cos{(2 \pi \beta i +\phi)}}{1-\lambda \alpha \cos{(2 \pi \beta i +\phi)}},\]
with $\lambda = V(t')^2/E^2$ and $\alpha\lambda = V/E$.

We therefore see that the subsystem $A$ of our model (with $t_B=0$) effectively behaves as a generalized Aubrey-Andr\'{e} (GAA) one-dimensional chain, which also exhibit similar ballistic, superdiffusive, and subdiffusive phases, as reported recently in Ref.~\cite{dynamics_2}.

\section{Conclusion}\label{sec:conclusions}
In summary, in this work, we study a two-leg ladder model, where one leg (called bath) is described by an AA-type Hamiltonian (which can undergo a localization-delocalization transition by tuning the strength of the incommensurate potential) and the other leg by the usual nearest-neighbor tight-binding model, which we refer to as the subsystem. By introducing the coupling $t'$ between these two legs, we obtain a rich phase diagram as shown schematically in Fig.~\ref{fig:phases}. We notice that beyond a critical coupling strength $t'_c$, there exists a regime where the subsystem can be completely localized. Also, we find different parameter regimes where the transport in the subsystem can be ballistic, subdiffusive, and superdiffusive in nature. However, it is important to note that there is an intermediate region, labeled as P4 phase in Fig.~\ref{fig:phases}, where the subsystem exhibits an anomalous phase structure. Here it may show super- or subdiffusive behavior based on parameter values and time scale of investigation. A more detailed study is required to gain better insight into the nature of phases in this crossover region.
We further explain some of these results using a toy model and also by mapping it to a well-studied 1D GAA model. 

Within the scope of this work, we could not study the nature of different phase transitions, associated scaling exponents and the universality classes. One of the reasons being that we do not have precise analytical (or even numerical) estimation for the phase boundaries. Further studies are required to address these issues. 

Our study has two primary implications: 1) it partially addresses the question of whether a localized bath can localize the system attached to it. At least in the non-interacting limit, our study finds that it is possible to have a suitable parameter regime where the bath can localize the system. 2) By doing this analysis, we also demonstrate 
a way to control the dynamics of a system (here, it is subsystem $A$) by not explicitly disturbing the system directly; instead, we have achieved that by altering another system (here, it is the bath $B$) and its coupling with $A$. This can have a significant application in quantum technologies and can potentially be implemented in experiments~\cite{exp1,exp2}.  

\begin{acknowledgments}
R.M. acknowledges the DST-Inspire research grant by the Department of Science and Technology, Government of India. 
\end{acknowledgments}


\appendix
\section{Results for random cosine model \label{cos_random}}
\begin{figure}[!ht]
    \centering
    \includegraphics[width=0.84\linewidth]{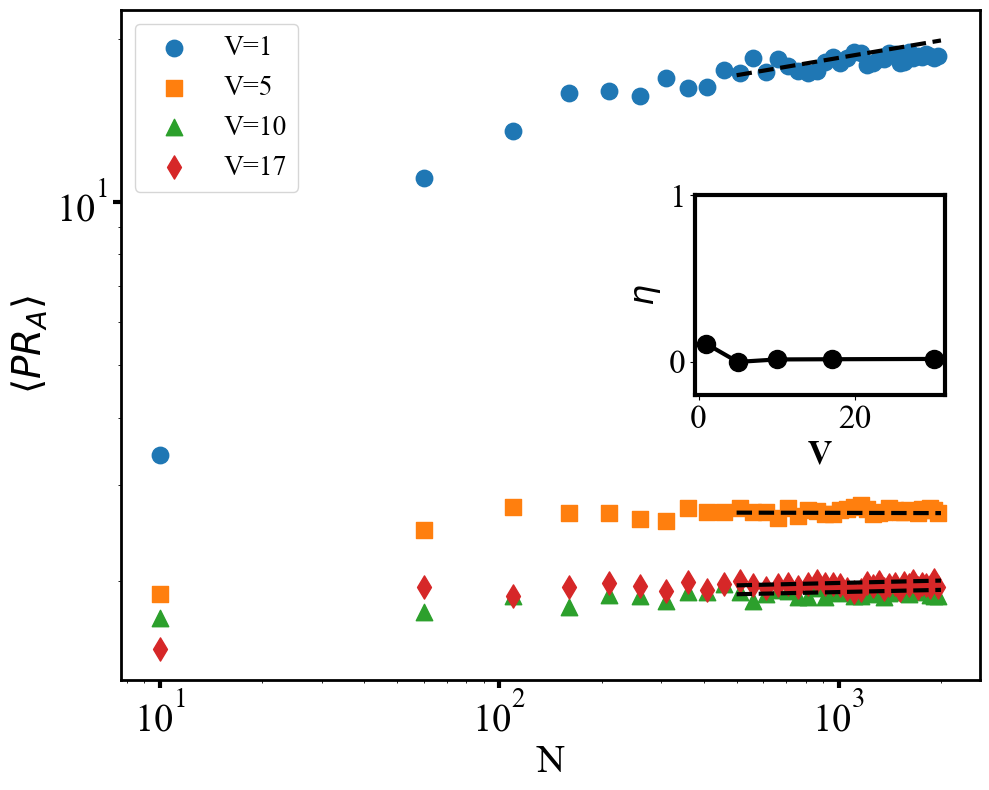}
    \caption{Log-log plot of disordered averaged $\langle PR_A \rangle$ vs N for $V=1,5,10,17$ ($\epsilon=0.5$ and $t'=5$). Inset: Scaling exponent ($\eta$) vs. $V$.}
    \label{fig:randm_phs}
\end{figure}

Instead of the correlated AA type disorder, if we take a random disorder in bath $B$, then the phase structure of the subsystem $A$ is expected to be completely different. To verify this, we replace the uniform phase factor $\phi$ in Eq. \ref{ham_eqn} by a random site-dependent phase $\phi_i$. We then calculate $\langle PR_A \rangle$ (averaged over multiple realizations of $\phi_i$) as function of the subsystem size ($N$). The results may be found in Fig. \ref{fig:randm_phs}. Fitting of the data by $\langle PR_A \rangle \sim N^{\eta}$ shows that the exponent ($\eta$) is very small regardless of the value of $V$. This indicates that a random (uncorrelated) potential in bath $B$ localizes all states in the subsystem $A$; this is in accordance with what is expected from the Anderson localization in one-dimensional system \cite{anderson.1958}.

\begin{figure}[]
    \centering
    \includegraphics[width=0.88\linewidth]{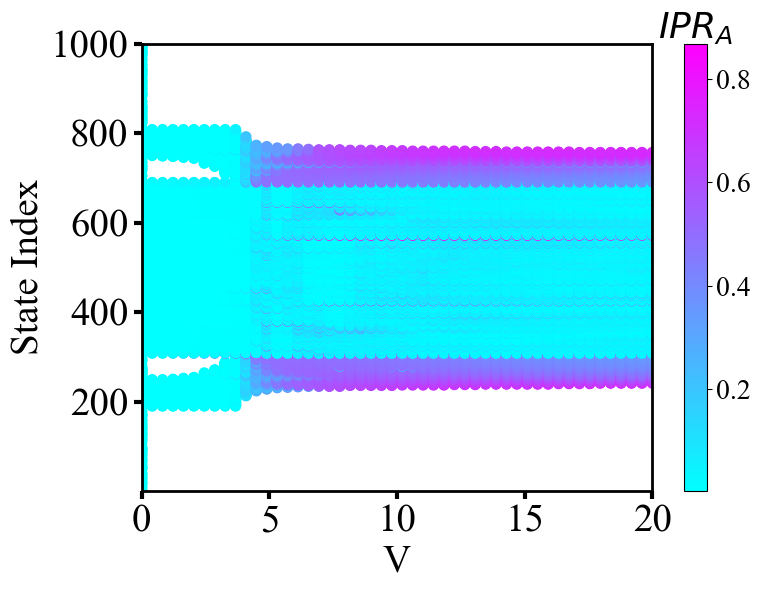}
    \caption{Variation of $IPR_A$ with $V$ for all $\epsilon$-significant states. Calculations are performed for the total subsystem size N = 500, $t'=3$ and $\epsilon =0.5$}
    \label{fig:IPR_tp_3}
\end{figure}

\begin{figure}[]
    \centering
    \includegraphics[width=0.84\linewidth]{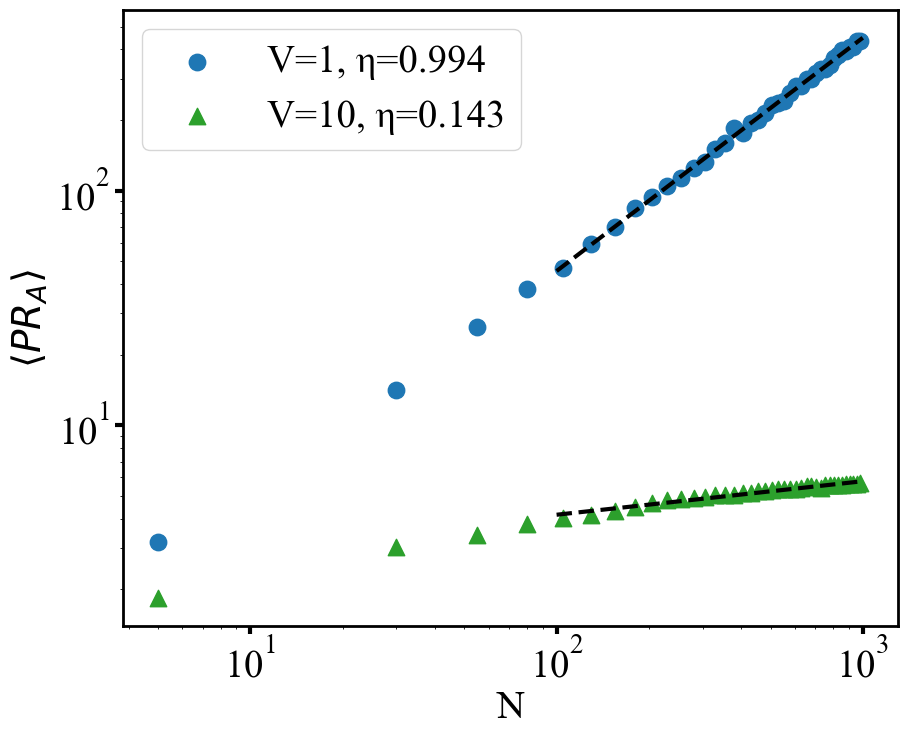}
    \caption{Log-Log plots for $\langle PR_A \rangle$ vs. $N$ with $t'=3$.  The scaling exponent $\eta$ is calculated for $V$ = 1 and $V$ = 10.}
    \label{fig:PR_tp_3}
\end{figure}

\section{Understanding Phases for $t'<t'_c$ \label{appendix_t'=3}} 
In this Appendix, we present results for $t'<t'_c \approx 4.4$. For definiteness, we focus on results corresponding to $t'=3$. The results for $t'/V\ll 1$ (P5 phase) will be discussed separately in Appendix \ref{very small t'}.

\begin{figure}[]
    \centering
\includegraphics[width=0.44\textwidth]{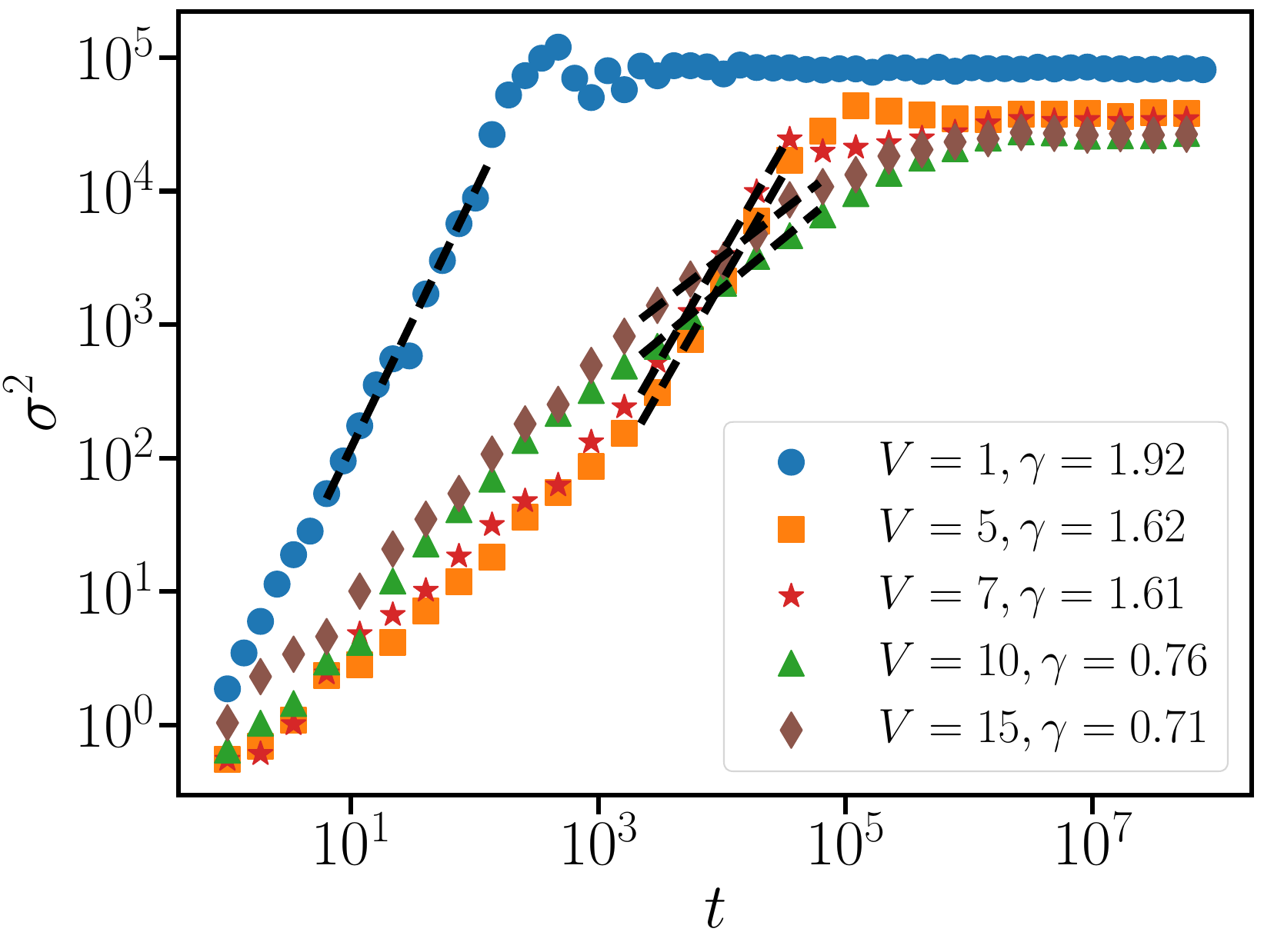}
\caption{$\sigma^2$ vs $t$ plot for different $V$, but for fixed $t'=3$ and  subsystem size $N=1000$. The black dashed line shows the fitting to the functional form $\sigma^2\sim t^\gamma$. }
    \label{app_dy}
\end{figure}

We have calculated $IPR_A$ for $ \epsilon =0.5$, and $ t' = 3$; we do not observe any localized region in the intermediate $V$ values. Corresponding results can be found in Fig. \ref{fig:IPR_tp_3}.  
We have further analyzed the scaling behavior of $\langle PR_A \rangle$ in different $V$ regions. The $\langle PR_A \rangle$ vs. $N$ plot (Fig.~\ref{fig:PR_tp_3}) suggests that at very low $V$, the subsystem A is completely delocalized with $\eta\approx1$, but if we increase $V$, we find some admixture of localized and delocalized states with $\eta < 1$. 

We have also investigated the dynamical properties of the subsystem for $t'=3$. Fig.~\ref{app_dy} shows for $V=1$, $\gamma\sim2$, which is a signature of ballistic growth; on the other hand, in this case for $V=10$ and $15$, we see a subdiffusive growth as the exponent $\gamma =0.76$.
In the intermediate regime, we also find a superdiffusive property (see $V=5$ and $V=7$ data), where $1<\gamma <2$. In this context, see our discussion in Appendix \ref{crossover regime} for more details.

\section{Energy-resolved IPR for subsystem, bath and full system \label{ER_IPR}} 
\begin{figure*}[t]
    \centering
    \includegraphics[width=15cm, height=11cm]{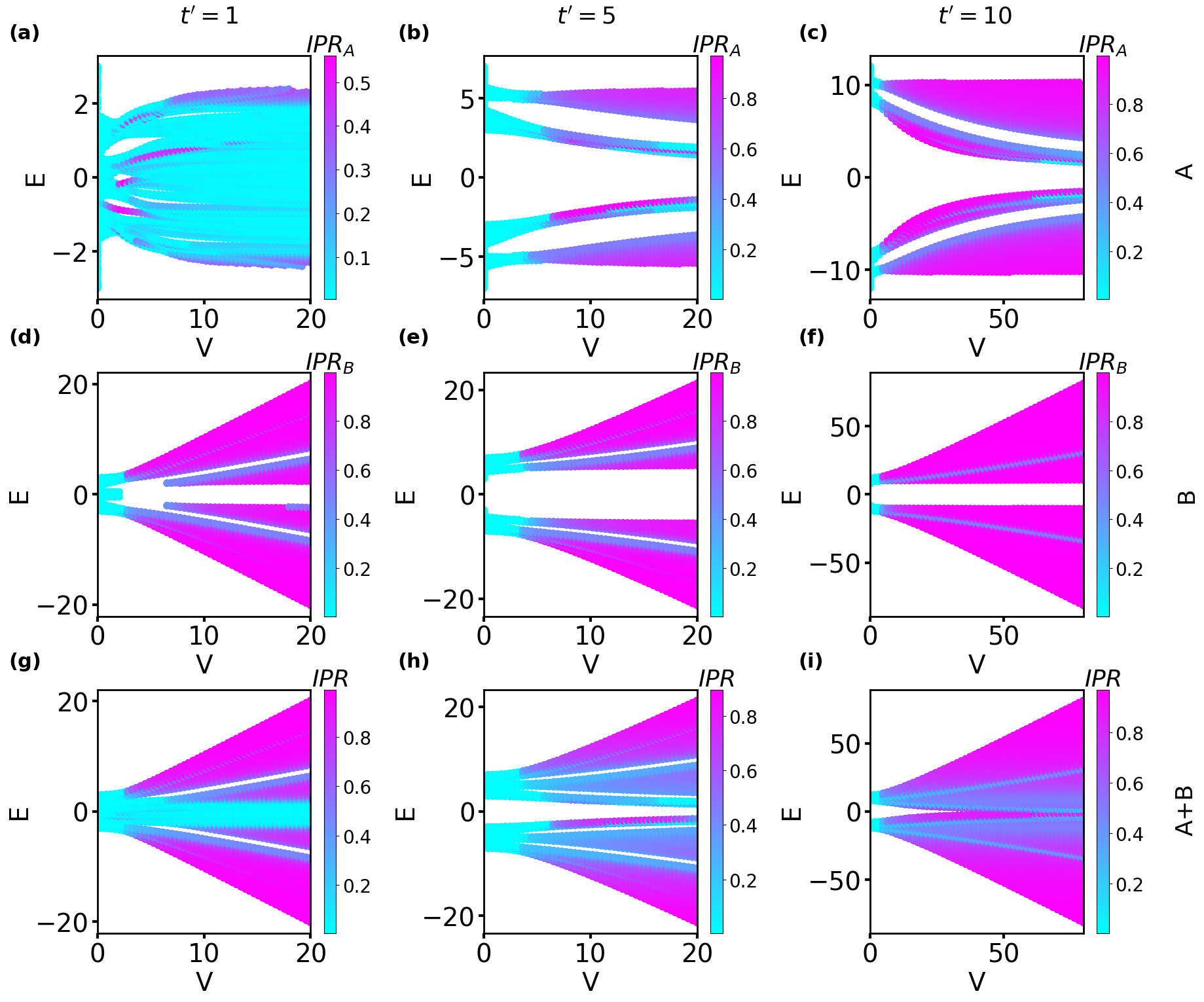}
    \caption{(a)-(c): Plots of energy-resolved $IPR_A$ as function of $V$ for subsystem $A$. (d)-(f): Plots of $IPR_B$ as function of $V$ for bath $B$. (g)-(i): Plots of energy-resolved $IPR$ as function of $V$ for full system $A+B$. Calculations are performed for the total subsystem size $N = 500$ and for $t'$ = 1, 5 and 10. For calculations of $IPR_A$ and $IPR_B$, we take only $\epsilon=0.5$ significant states.}
    \label{ER_IPR_V}
\end{figure*}

To investigate possible correlations between the energy scales of the states and their localization properties, we plot the energy-resolved IPR for subsystem $A$, bath $B$, and the full system $A+B$. For subsystem $A$, we compute $IPR_A$ as defined in Eq.~\ref{IPR_eq}; analogously, we evaluate $IPR_B$ for the bath. For the full system, we calculate the IPR corresponding to its actual eigenkets (rather than their projections). The corresponding results are presented in Fig.~\ref{ER_IPR_V}.

Our results for a fixed system size suggest that the full system, $A+B$, can exhibit some form of mobility edge. On the other hand, the concept of the mobility edge is ill-defined for the subsystem - since the projected states $\ket{\psi_n}$ are not the energy eigenstates of its Hamiltonian $H_A$. 
For our study of subsystem localization, a more useful quantity is the fraction of significant states that are delocalized (weakly delocalized). In the following we discuss this in details.

In analyzing the localization phase diagram of the subsystem in the thermodynamic limit, the concept of a mobility edge does not play particularly any important role. In fact, the scaling relation, $\langle PR_A \rangle \sim N^{\eta}$, remains valid even when the subsystem shows mobility edges. To understand this, we write $\langle PR_A \rangle$ as a sum of two different averages:
\begin{equation}
    \langle PR_A \rangle = f_D \langle PR_A \rangle_D + f_L \langle PR_A \rangle_L,
\end{equation}
where $\langle PR_A \rangle_D$ is the $PR_A$ averaged over delocalized or weakly delocalized $\epsilon$-significant states, and similarly, $\langle PR_A \rangle_D$ is the $PR_A$ averaged over localized $\epsilon$-significant states. In the equation, $f_D=|S_{\epsilon,D}|/|\mathcal{S}_{\epsilon}|$ and $f_L=|S_{\epsilon,L}|/|\mathcal{S}_{\epsilon}|$. We recall that $\mathcal{S}_{\epsilon}$ is the set of $\epsilon$-significant states and $|\mathcal{S}_{\epsilon}|$ denotes the number of elements in the set. Similarly, $S_{\epsilon,D}$ and $S_{\epsilon,L}$ represent, respectively, the set of delocalized (including weakly delocalized) states and the set of localized states. The number of elements in them are, respectively, represented by $|S_{\epsilon,D}|$  and $|S_{\epsilon,L}|$. Since $|S_{\epsilon}|=|S_{\epsilon,D}|+|S_{\epsilon,L}|$, we have 
\begin{equation}\label{fd_fl}
    f_D+f_L=1,
\end{equation}
with $f_D, f_L\ge 0$.

In the thermodynamic limit, $\langle PR_A \rangle_L=O(1)$ and $\langle PR_A \rangle_D=O(N^{\eta})$, where $0<\eta\le 1$. Now, due to the condition in Eq. \ref{fd_fl}, we can have the following three situations in the thermodynamic limits: (a) $f_D=0$ and $f_L=1$, (b) $f_D=1$ and $f_L=0$, and (c) $0<f_D,f_L<1$. In the first and second cases, the subsystem will be in the localized and delocalized (weakly delocalized) phases, respectively. The third case is an interesting case where the subsystem will have both localized and delocalized (weakly delocalized) states. This case also accounts for the situation where the subsystem has mobility edges.  
In the last two cases (i.e., whenever $f_D>0$), $\langle PR_A \rangle \sim N^{\eta}$, as we hypothesized at the beginning. From the log-log plot of $\langle PR_A \rangle$ vs $N$, one can extract the exponent $\eta$ and also estimate the fraction of delocalized (including weakly delocalized) states $f_D$. The value of $\eta$ helps determine if the subsystem is in delocalized, localized, or in weakly delocalized (fractal) phase.

\section{A measure for bond-localization \label{bond_local}}
A general eigenket of the full system is given by
\begin{equation}\label{eq_ket}
\ket{\xi}=\sum_{i=1}^N\left(a_{i}\ket{i}+a_{N+i}\ket{N+i}\right),
\end{equation}
where $\sum_{i=1}^N \left(|a_i|^2+|a_{N+i}|^2\right)=1$.
We recall that the $i$th site in the subsystem $A$ is directly connected to the $(N+i)$th site in the bath $B$ by the hopping parameter $t'$. A straight forward calculation of the bond localization parameter (as defined in Eq. \ref{kappa_para}) shows that
\begin{equation}
\kappa_\xi = 4\sum_{i=1}^N \left[Re(a_i^*a_{N+i})\right]^2.
\end{equation}
It is not difficult to see that the value of $\kappa_\xi$ is bounded by 1. This can be shown in the following way.
\begin{equation}
\begin{split}
    \kappa_\xi & \le 4 \sum_{i=1}^N |a_i^*a_{N+i}|^2  \\
               & = 4 \sum_{i=1}^N |a_i|^2 |a_{N+i}|^2   \\
               & \le 4(\sum_{i=1}^N |a_i|^2)(\sum_{i=1}^N |a_{N+i}|^2)\\
               & \le 4\left[\frac 12 \left(\sum_{i=1}^N |a_i|^2+\sum_{i=1}^N |a_{N+i}|^2\right)\right]^2\\
               &=1.
\end{split}
\end{equation}
In the last line, we have used the normalization condition, $\sum_{i=1}^N \left(|a_i|^2+|a_{N+i}|^2\right)=1$.

We now determine the values of $\kappa_\xi$ for different types of states.

\begin{enumerate}
\item Dimer state: consider an ideal dimer state $\ket{\xi}=a_i\ket{i}-a_{N+i}\ket{N+i}$ for some $i$. We take $a_i=a_{N+i}=\frac{1}{\sqrt{2}}$ and $a_j=0$ for $j\ne i, N+i$. For this state, $\kappa_\xi = 1$.
\item Delocalized state in subsystem $A$:  Consider a state that is fully delocalized in $A$ but does not have presence in $B$. For this state $|a_i|=\frac{1}{\sqrt{N}}$ and $a_{N+i}=0$ for all $i$ in Eq. \ref{eq_ket}. For this state, $\kappa_\xi =0$.
\item Delocalized state in full system: consider a state that is fully delocalized over the full system. For this state $|a_i|=|a_{N+i}|=\frac{1}{\sqrt{2N}}$ for any $i$ in Eq. \ref{eq_ket}. For this state, $\kappa_\xi=\frac{1}{N}$, which vanishes in the thermodynamic limit.
\end{enumerate}

\begin{figure}[]
    \centering
\includegraphics[width=0.44\textwidth]{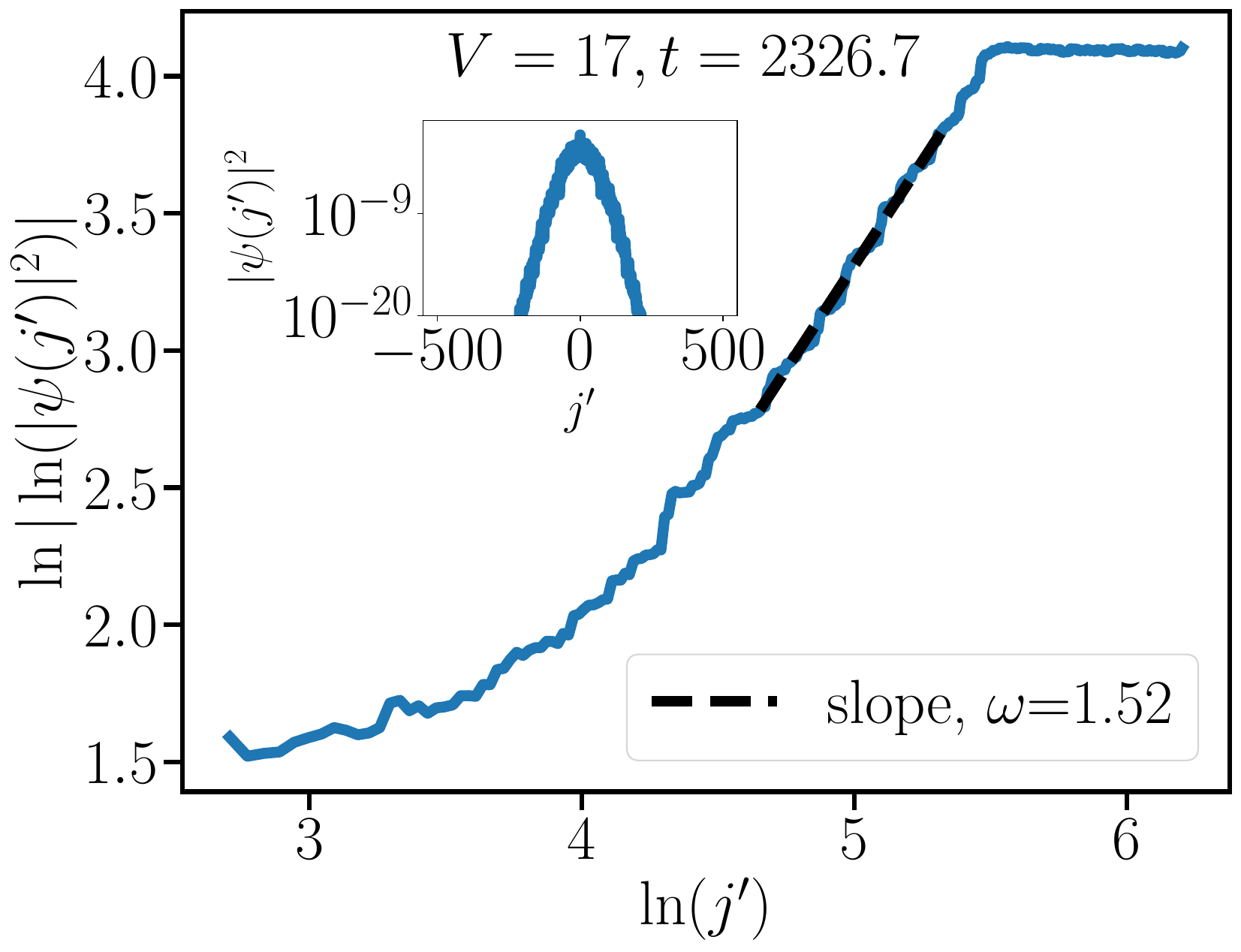}
\caption{$\ln|\ln(|\psi(j')|^2)|$ vs. $\ln(j')$ plot for $j'>0$ for $V=17$ and $N=1000$. The dashed line corresponds to the best fit with slope $\omega=1.52$. The inset shows the probability distribution $|\psi(j')|^2$ vs. $j'$, where $j'=j-N/2$.}
    \label{Fig_exp}
\end{figure}

\begin{figure}[]
    \centering
\includegraphics[width=0.46\textwidth]{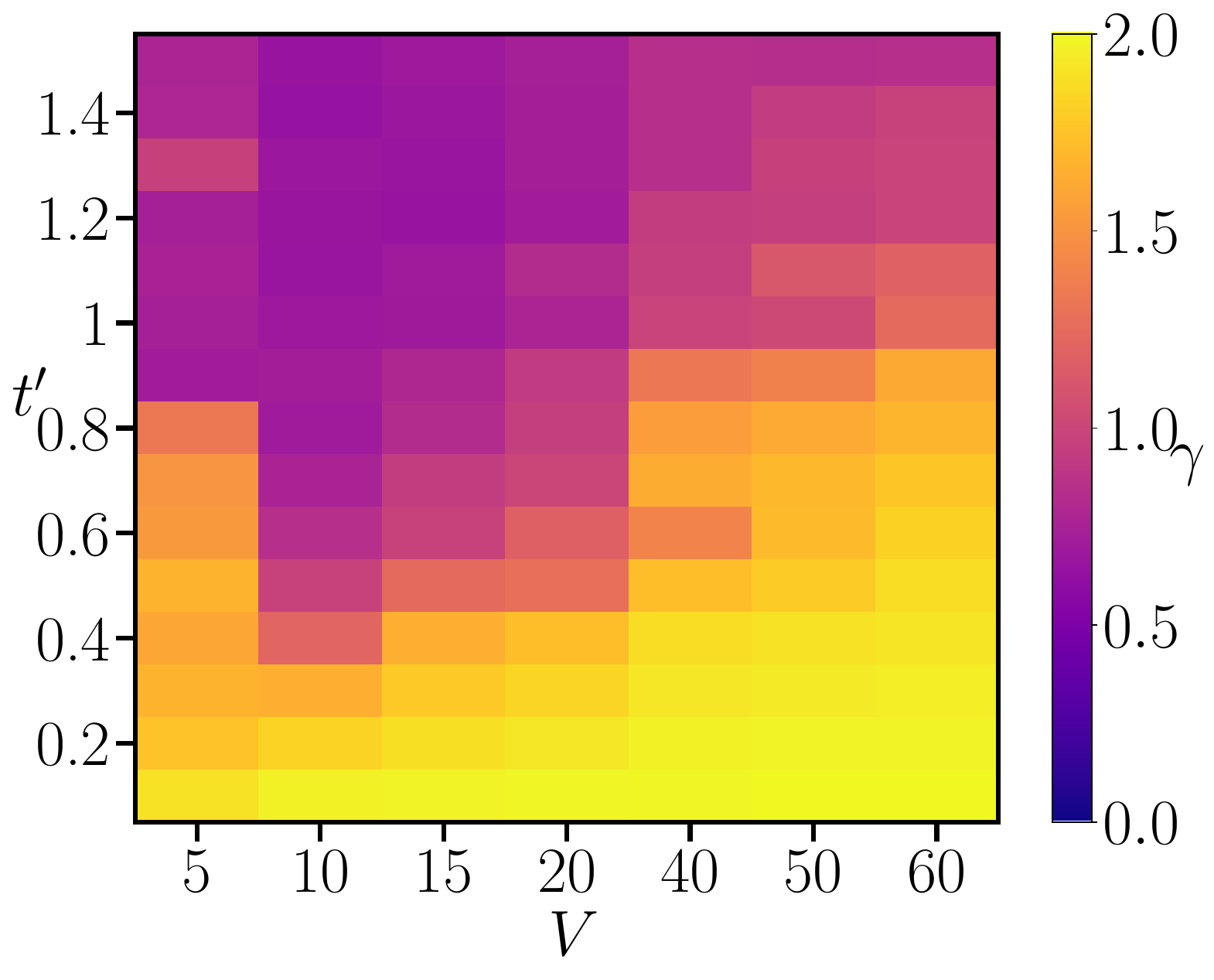}
\caption{Contour plot of the scaling exponent $\gamma$ as a function of $t'$ and $V$, where we restrict ourselves to small $t'$ regime. }
\label{Fig_contour_smallt}
\end{figure}

\begin{figure}[]
    \centering
\includegraphics[width=0.48\textwidth]{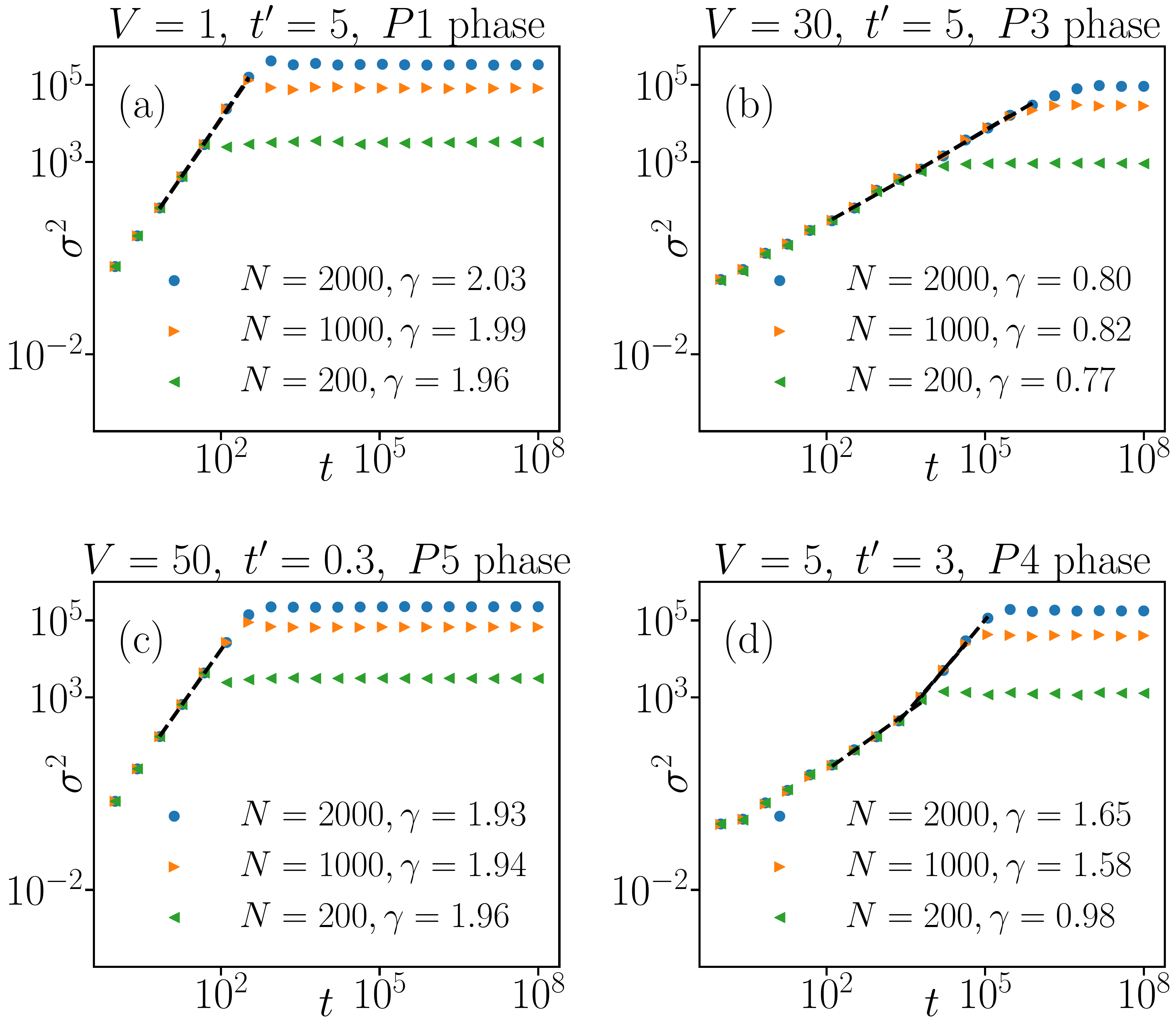}
\caption{Show the system size dependence on the exponent $\gamma$ for different phase.}
\label{system_size_gamma}
\end{figure}

\begin{figure}[]
    \centering
\includegraphics[width=0.48\textwidth]{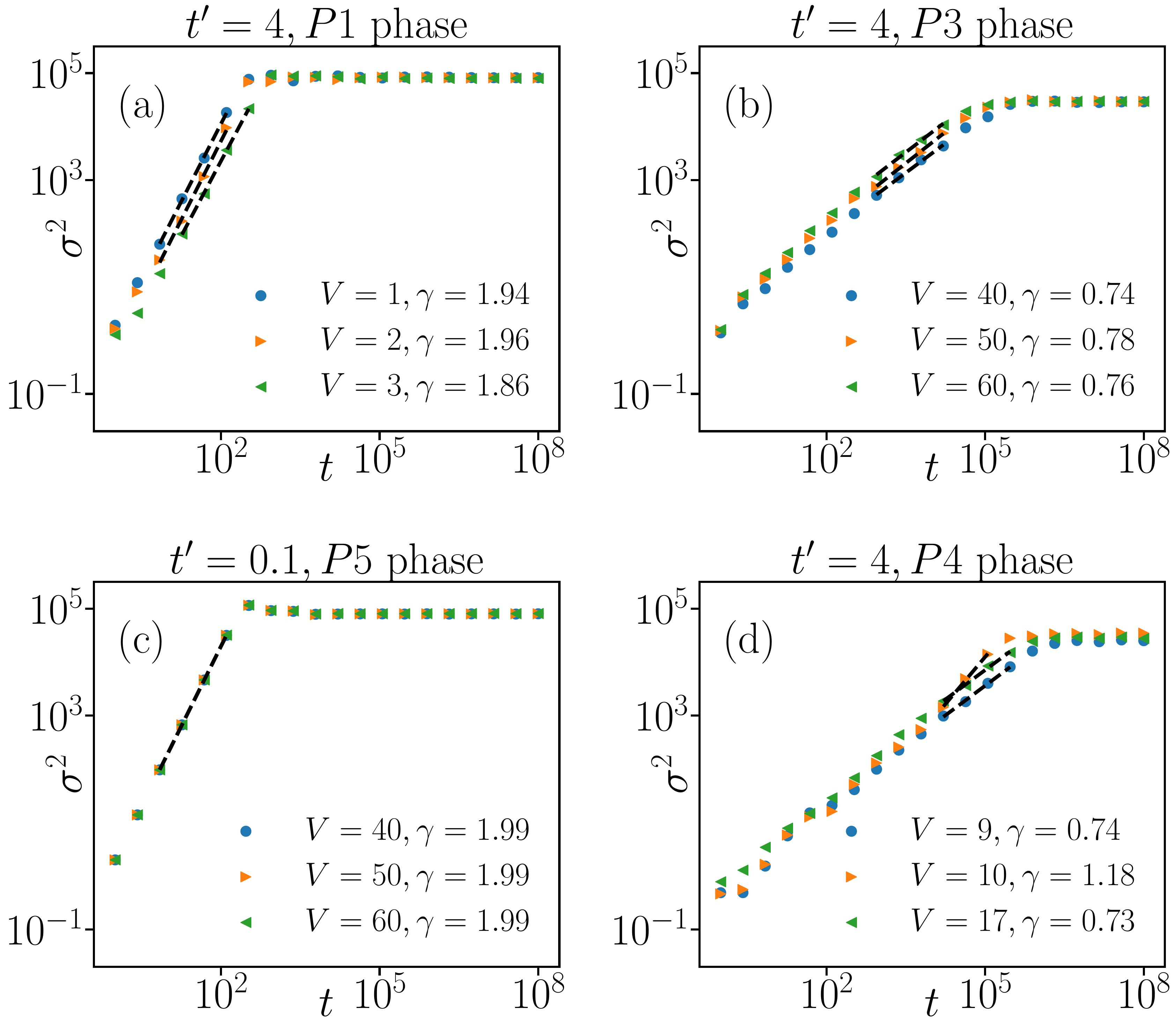}
\caption{Show the variation of the  exponent $\gamma$ with $V$ in each different phases.}
\label{unstable_gamma_p4}
\end{figure}

\section{Scaling of projected probability distribution \label{appendix_psi_scaling}}

It has been argued recently in Ref.~\cite{streched_exp} that if the transport is sub (super) diffusive, the probability distribution at large time will have a stretched-exponential form.  Moreover, that stretched exponent is related to the scaling exponent $\gamma$; note $\sigma^2\sim t^{\gamma}$. 
In our context, if one starts from the same initial state $|\phi_0\rangle=|1_{N/2}\rangle$ as mentioned in the main text, the probability distribution corresponding to the projected state in A obeys the following scaling form in the large $t$ and $j'=j-N/2$ limit, 
\begin{equation}
 |\psi(j',t)|^2\sim a(t)e^{-b(t)|j'|^{\omega}}
\end{equation}
where $a(t)$ and $b(t)$ are functions of $t$, and $\omega =\frac{2}{2-\gamma}$ (see Ref.~\cite{streched_exp}).  Here, we obtain the scaling exponent $\omega=1.52$ by fitting  $|\psi(j',t)|^2$ with $j'$ for $V=17$ and for large enough $t$ (see Fig.~\ref{Fig_exp}).  It implies $\gamma\approx 0.7 <1$; sub-diffusive transport, precisely what we have observed in the main text.

\section{P5 phase: A contour plot for scaling exponent of $\gamma$ \label{very small t'}}
In Fig.~\ref{Fig_contour_smallt}, we present a zoomed-in view of the small-$t'/V \ll 1$ regime of Fig.~\ref{Fig5_1} (the contour plot of $\gamma$ in the $t'$--$V$ plane). This figure clearly shows a crossover from super-diffusive to ballistic behavior in the small-$t'$ regime. We recall that the subsystem $A$ shows ballistic behavior in $t'\to 0$ limit. We identify this as P5 phase. The extent of this P5 phase grows with increasing $V$, suggesting that in the $V \to \infty$ limit, the system approaches the ballistic phase for any finite $t'$. This behavior is intuitively expected: in this limit, the coupling between subsystem $A$ and bath $B$ becomes irrelevant, and the dynamics is effectively governed solely by $H_A$ provided we initiate the dynamics in the subsystem $A$.

\section{P4 phase: Strong dependence on system size and model parameters  \label{crossover regime}} 
Here, we discuss the strong system-size and model parameter dependence of the dynamical scaling exponent $\gamma$. Figure~\ref{system_size_gamma} shows the variation of $\sigma^2$ with time $t$ for different system sizes across various phases. We find that the exponent $\gamma$ remains essentially unchanged for $N=200$, $1000$, and $2000$ in the $P1$ (ballistic), $P3$ (sub-diffusive), and $P5$ (ballistic/super-diffusive) phases. However, in the $P4$ phase, the variation of $\gamma$ with system size is significant. In fact, as shown in Fig.~\ref{system_size_gamma}, the exponent extracted for $N=200$ suggests sub-diffusive behavior, whereas for $N=1000$ and $2000$ it indicates super-diffusive behavior for the same choice of parameters within the $P4$ regime. This makes it extremely difficult to unambiguously determine the dynamical character of this phase.

Moreover, within a stable phase, one expects the system to exhibit similar behavior in the neighborhood of any given point in parameter space. This is indeed what we observe in Fig.~\ref{unstable_gamma_p4} for the P1 (ballistic), P3 (sub-diffusive), and P5 (ballistic/super-diffusive) phases, where all nearby points display consistent dynamical exponents. In contrast, within the P4 phase, we find non-monotonic behavior of $\gamma$ with respect to a small change in a parameter. To illustrate this more clearly, we plot $\sigma^2$ as a function of $t$ for $V=9$, $10$, and $17$ with $N=1000$. Interestingly, the dynamics are super-diffusive for $V=10$, while in the nearby points, like $V=9$, it shows sub-diffusive behavior. Such behavior is not expected if the system were deep inside a stable phase. Together with the strong system-size dependence of $\gamma$ in the P4 regime (see Fig.~\ref{system_size_gamma}), these observations suggest that the P4 phase reported in Fig.~\ref{fig:phases} is more likely an anomalous crossover regime that needs further study, which lies beyond the scope of the present work.
\nocite{*}

\bibliography{manuscript}

\end{document}